\documentclass[epj]{svjour}
\usepackage{graphics,natbib}
\usepackage{xcolor}
\usepackage{graphicx}

\begin{document}
\title{Understanding the Engines and Progenitors of Gamma-Ray Bursts}
\subtitle{}
\titlerunning{Probing GRBs}
\authorrunning{Fryer et al.}
\author{Chris L. Fryer\inst{1,2}, Nicole Lloyd-Ronning\inst{1,3}, Ryan Wollaeger\inst{1}, Brandon Wiggins\inst{1,4}, Jonah Miller\inst{1,5},  Josh Dolence\inst{1}, Ben Ryan\inst{1}, Carl E. Fields\inst{1,6}% etc
% \thanks is optional - remove next line if not needed
%\thanks{\emph{Present address:} Insert the address here if needed}%
}                     % Do not remove
\institute{Center for Theoretical Astrophysics, Los Alamos National Laboratory, Los Alamos, NM 87544
\and
Department of Physics, The George Washington University, Washington, DC 20052
\and
University of New Mexico, Los Alamos, NM 87544
\and
Department of Physical Science, Southern Utah University, Cedar City, Utah 84721
\and
Center for Nonlinear Studies, Los Alamos National Laboratory, Los Alamos, NM, USA
\and
Department of Physics and Astronomy, Michigan State University, East Lansing, MI 48824}
\date{Received: date / Revised version: date}
% The correct dates will be entered by Springer
%
\abstract{
Our understanding  of the engines and progenitors of gamma-ray bursts has expanded through the ages as a broader set of diagnostics has allowed us to test our understanding of these objects.  Here we review the history of the growth in our understanding, focusing on 3 leading engines and 9 potential progenitors.  The first gravitational wave detection of a short burst is the latest in a series of breakthrough observations shaping this understanding and we study the importance of multi-diagnostic, multi-messenger observations on these engines and their progenitors.  Our understanding based on a detailed study of nearby bursts can be applied to make predictions for the trends expected as we begin to observe high redshift bursts and we discuss these trends.  
 % end of PACS codes
} %end of abstract
\maketitle
\section{Introduction}
\label{intro}

50 years ago, the Vela satellites observed the first in a series of bursts of gamma-ray emission.  The first 16 bursts were published in a compilation in 1973~\citep{klebesadel73}.  The road to understanding
the nature (energies, populations, engines, progenitors, environments, etc.) of these gamma-ray bursts (GRBs) has sometimes wandered, but a broad range of  diagnostic signposts have led to a continual refinement of GRB models.  In this paper, we review this 50 year journey in understanding the origin of,
and engines behind, these powerful cosmic explosions.  Observations have whittled down the hundreds of models arising from the creative minds of theorists to a few classes of models. These varied and multi-messenger constraints have shaped our understanding of the origin of these
bursts.  The detection of gravitational waves associated with a gamma-ray burst is just one (albeit a very important one) in a long history of diagnostics used in this study.  The many diagnostics
(using a broad range of photon energies probing different aspects of the explosions as well as gravitational waves) have helped, and continue to help, astrophysicists probe the details of the engines and
progenitors of these bursts, in addition to studying the fundamental physics behind these engines.  In this paper, we review the wide range
of diagnostics that have shaped our understanding of the short-duration GRB (sGRB) engine.  We study these diagnostics over the broad 50 year journey, beginning with the observational 
constraints from the first 3 decades (Section~\ref{sec:energetics}) and the models (Section~\ref{sec:engine}) and progenitors (Section~\ref{sec:prog}) 
favored by these observations.  The distribution of durations (Section~\ref{sec:dur}) allowed astrophysicists to distinguish between progenitor models and more detailed calculations led to a set of observational predictions (Section~\ref{sec:pred}).  Observations confirming some of these predictions 
have led to the emergence of a favored engine behind GRBs.  But, as we have seen with GW170817, the iterations between theory and observation have only just begun with better data making more detailed theory necessary.  We conclude with a description of future prospects with GRB observations.

\section{Observations and Models in the first 3 decades}
\label{sec:energetics}

With the discovery of GRBs, astrophysicists began to design missions 
focused on finding more.  The Burst And Transient Source Experiment
(BATSE)~\citep{fishman93} produced a large database of gamma-ray
bursts, allowing astrophysicists to discern different GRB subgroups:
short, hard bursts and long, soft bursts~\citep{mazets83,kouveliotou93}.  The
short bursts had durations from a few milliseconds to a few seconds.
The long bursts were typically a few seconds to a few hundred seconds in duration although classes of very and ultra-long bursts exist with timescales exceeding 1000\,s~\citep{levan14}.  There were spectral differences as well: short bursts typically had higher hardness ratios (more flux in higher
energy bands) than the long bursts.  In both cases, the gamma-ray
emission is highly variable with variability timescales below 1\,s.

Models for these bursts ranged from nearby events (lightning storms in the upper Earth's atmosphere) to very distant explosions (cosmic string interactions in the early universe)~\citep{nemiroff94}.  Initially, all models were possible.  But the growing observational sample began to constrain these proposed models.  For example, although the gamma-ray observations did not exactly pinpoint the burst location, astrophysicists were able to statistically study their
distribution, finding that these bursts were isotropically
distributed~\citep{briggs93}.  Most Earth atmosphere or Galactic models struggled
to reproduce these observations.  Without a probe beyond the
gamma-rays, these arguments were limited; e.g., astrophysicists could 
test distances by imposing specific theoretical
models to make predictions on the distributions.  This changed when the accurate
localization of GRB970228 by the Italian-Dutch BeppoSAX satellite led to the discovery of its X-ray and optical afterglow and subsequent association with a host galaxy~\citep{costa97,vanparadijs97}.  By assuming the association of the burst with a host galaxy was not coincidence, astrophysicists 
could pin down the distance of the burst.  GRB970508 brought even more conclusive evidence that these bursts must be extragalactic.  In the follow-up optical  measurements of this burst, astrophysicists observed absorption lines
indicating a redshift of at least 0.835~\citep{metzger97}.  Radio
scintillation measurements that, coupled with the assumption of a
nearly speed of light expansion, confirmed the spectral redshift
measurement~\citep{frail97}.  After these two bursts, there was no 
doubt that some bursts were extragalactic and, with these large distances, astrophysicists could place energy constraints on the bursts and these high energies,
$\sim 10^{51}\,{\rm erg}$ (after corrected for beaming), required catastrophic events. Let us review the diagnostics in the first 30 years:
\begin{itemize}
\item{The gamma-rays from GRBs are variable on the millisecond timescale}
\item{The duration of the gamma-ray emission ranges from a few
  milliseconds to few hundred seconds}
\item{Two (at least two) classes of bursts exist (short, hard and long,soft) and it may be that
  multiple models are required to match the data}
\item{Bursts require supernova-like energetics ($10^{51}\,{\rm erg}$)}
\item{The broadband spectra appear to be power-laws without features and a peak near an MeV.}
\end{itemize}
These constraints, obtained by diagnostics that spanned a range of
wavelengths and probed different aspects of the explosion, were
sufficient to drive the development of the current ``standard'' GRB
models.  In the next few sections, we review these models and discuss
how the multi-probe diagnostics were used to focus on the GRB 
engine.

\section{Engines and Energetics}
\label{sec:engine}

The rapid variability of GRBs suggests extreme conditions where the emission environment changes on millisecond timescales.  For the powerful engines needed to explain extragalactic sources, the options for rapid variability quickly narrows down to only a few possibilities.  Engines active just above the surface of a neutron star (NS) or black hole (BH) soon became favored.  The timescale for variabilities in the accretion disk on a NS or BH or instabilities in the magnetic field structure or spin down in a rapidly rotating NS will be on order of this sound crossing or orbital time, e.g.:
\begin{equation}
  t_{\rm variability} \approx R_{\rm NS}/c_s \approx 0.1 {\rm ms}
\end{equation}
where $R_{\rm NS} \approx 10^6 {\rm \, cm}$ and $c_s \approx 10^{10}
{\rm \, cm \, s^{-1}}$.  The potential to produce $10^{51} {\rm \, erg}$ bursts 
coupled to the ability to have rapid variability led astrophysicists to focus on these NS and BH engines.  The difficulty then became determining how to drive an 
engine on these compact objects.  A number 
of power sources have been suggested, for example:
\begin{itemize}
\item{Neutron Star Phase Transition:  Burst powered by the potential energy release when a phase transition to quark matter occurs in a neutron star~\citep{1980Natur.287..122R,1996ApJ...462L..63M}}
\item{Magnetar:  Burst powered by a rapidly spinning neutron star with 
strong magnetic fields~\citep{duncan96}}
\item{Neutron star accretion disk (NSAD): Burst powered by accretion onto a neutron star}
\item{Black hole accretion disk (BHAD): Burst powered by accretion
  onto a black hole~\citep{narayan92,woosley93}}
\end{itemize}

This is far from a complete list, but it gives a flavor of these compact remnant models.  Simulations of phase transitions have not produced the high Lorentz factors needed, e.g., see ~\cite{1998ApJ...501..780F}.  We will thus consider only the latter 3 models, first studying the magnetar engine.  With strong (magnetar-strength: $\sim 10^{15} {\rm
  \,Gauss}$) magnetic fields, a pulsar can quickly release the
rotational energy of a newly formed neutron star.  To calculate the
energy available for such a model, we need to estimate the rotational
energy of the neutron star.  The moment of inertia of neutron stars
depends upon the equation of state~\citep{worley08}, but all estimates
of the moment for a neutron star ($I_{\rm NS}$) are within a factor of
two of:
\begin{equation}
I_{\rm NS} = 10^{45} (M_{\rm NS}/M_\odot) {\rm g \, cm^2}
\end{equation}
where $M_{\rm NS}$ is the neutron star mass.  The corresponding 
rotational energy ($E_{\rm rot}$) is:
\begin{equation}
E_{\rm rot} = 1/2 I_{\rm NS} \omega^2 = 5\times10^{50} (\omega/1000Hz)^2 \, {\rm erg}
\end{equation}
where $\omega$ is the angular velocity.  If the neutron star is spinning with 
a ms period, it can produce a $10^{51}$\,erg explosion if it can tap 10\% of 
the rotational energy to drive a jet.

The energy released in an accreting neutron star or black hole is set by the 
potential energy released by the accreting matter ($E_{\rm acc}$):
\begin{equation}
E_{\rm acc} = G M_{\rm NS,BH} m_{\rm acc}/R_{\rm NS,BH} - G M_{\rm NS,BH} m_{\rm acc}/R_0
\end{equation}
where $G$ is the gravitational constant, $M_{\rm NS,BH}$ is the
neutron star or black hole mass, $m_{\rm acc}$ is the accreted mass,
$R_{\rm NS,BH}$ is the neutron star radius or the black hole's
innermost stable circular orbit (ISCO), and $R_0$ is the initial
radius of the infalling material.  Typically the initial radius is sufficiently high to make this right term negligible.  For the neutron star, the energy released 
is roughly:
\begin{equation}
E_{\rm acc} = 3.7 \times 10^{51} (m_{\rm acc}/0.01 M_\odot).
\end{equation}
If the engine is 3\% efficient at converting this energy into jet
energy, accreting 1/10th of a solar mass would be sufficient to drive
a $10^{51} \, {\rm erg}$ explosion.  But to extract this energy and
drive a jet, an accretion disk must form around the neutron star to wind up the magnetic fields.  The BH accretion disk model releases more potential energy than the NS model, but extracting the energy might be more difficult.  For these BHAD models, the disk is required to prevent the material from flowing directly into the black hole.  Although accreting NSs need not form a disk to produce a jet, most engines require the formation of an accretion disk to produce the strong magnetic fields believed to produce relativistic jets.  

For all three models, the formation of GRB jets is based on analogies
with other astrophysical phenomena and not direct simulations.  For
example, although no simulations exist that produce
highly-relativistic jets from magnetars (indeed, there is a known
issue of baryon loading for magnetar jets in newly formed neutron
stars~\cite{2014ApJ...788L...8M}), we know that normal pulsars produce
energetic jets and it is plausible to assume that the higher-power
jets in magnetars will have higher Lorentz factors.  Similarly,
current magnetohydrodynamic calculations have produced jets and we
know that jets are produced in active galactic nuclei accretion disks.
These calculations push the limit of current capabilities and it is
not surprising that no calculation has captured this physics
completely. However, it is plausible to assume that the more extreme
accretion scenarios discussed for GRB progenitors will produce
higher-Lorentz factor jets.

\begin{figure}
% Use the relevant command for your figure-insertion program
% to insert the figure file.
% For example, with the option graphics use
\resizebox{0.48\textwidth}{!}{%
  \includegraphics{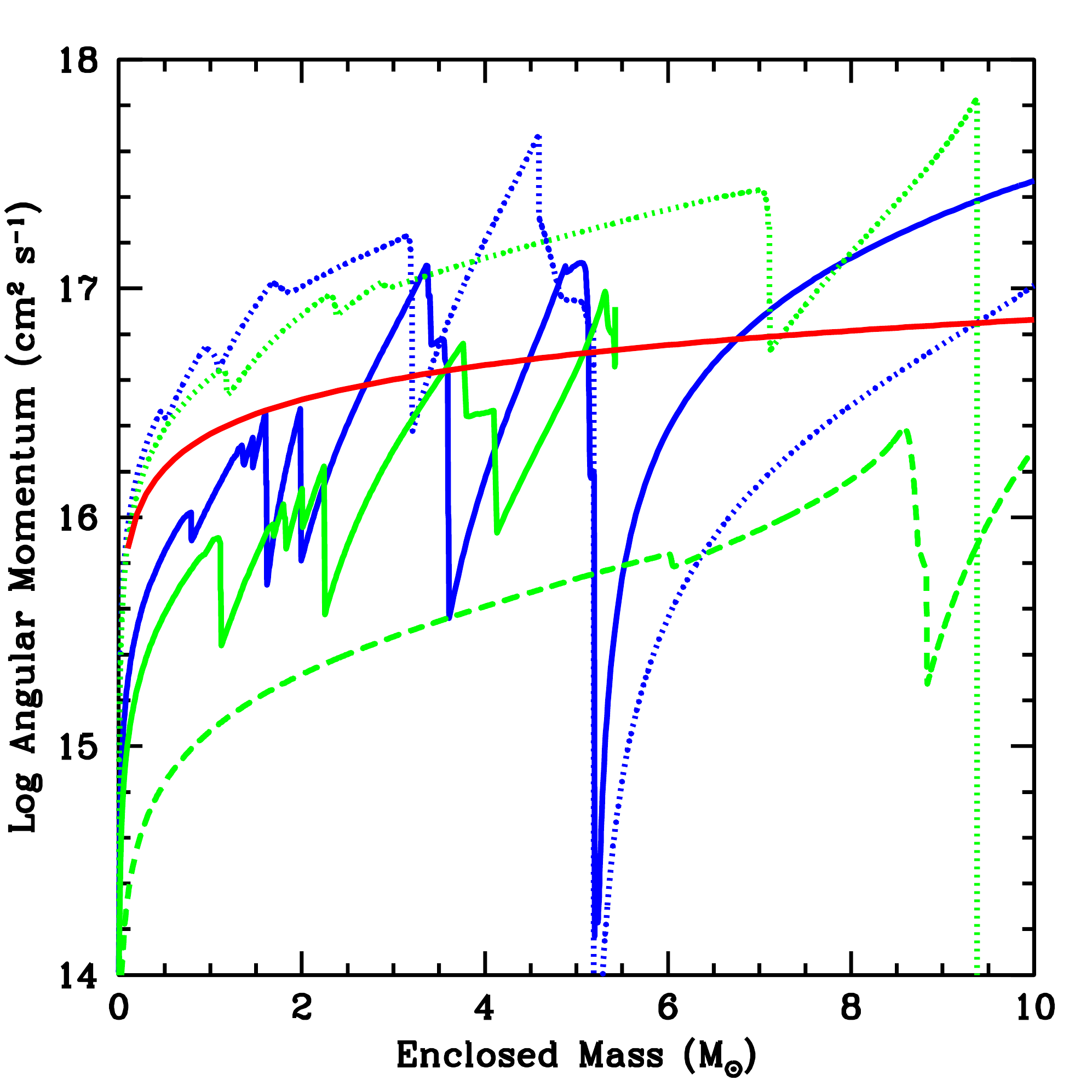}
}
\caption{Angular momentum profile (specific angular momentum versus of
  enclosed mass) for two different stellar models:
  $15\,M_\odot$(blue), $25\,M_\odot$ (green).  The dotted lines
  correspond to models using the GENEC code with no magnetic
  dynamo~\citep{belczynski18}, the solid lines correspond to models
  using the Kepler code using a weak magnetic dynamo~\citep{heger05}, and the
  dashed line corresponds to simulation using a stronger magnetic
  dynamo using the MESA code~\citep{paxton13}.  The solid red line corresponds 
  to the angular momentum needed to make a 1000\,km disk around the compact 
  remnant.}
\label{fig:jcomp} 
\end{figure}

All three models successfully explain the energetics if the engine has sufficient angular momentum.  Different progenitors
produce different angular momentum profiles and we will discuss this
in detail in Section~\ref{sec:prog}.  But to get a flavor of the
issues tying energetics to our engines, let us review single star
models and the kinds of angular momenta they produce.
Figure~\ref{fig:jcomp} shows the angular momentum profile (specific
angular momentum versus enclosed mass) for stellar-evolution models
using 3 different prescriptions for the magnetic dynamo developed
between burning layers in the star.  On top of these profiles is 
plotted the angular momentum needed to form a disk
at 1000\,km.  The angular momentum is lowest in the center of the star.
The fact that the angular momentum increases with mass coordinate led
scientists~\citep{woosley93,fryer99a} to argue for black hole accretion
disk models over neutron star models.  For the stellar models without
any dynamo (dotted lines), the specific angular momentum is high,
forming a disk whether the compact remnant formed is a neutron star or
black hole.  The neutron star spin would be less than a
sub-millisecond and a potential magnetar model would be be able to tap
$10^{52} {\rm \, erg}$ of rotational energy.  However, with even a
weak dynamo (solid lines), a disk only forms once the compact remnant
exceeds 3\,M$_\odot$.  Such stars could not produce a NSAD engine.
For the magnetar model, these stars would produce a millisecond
neutron star, requiring efficiencies in excess of 25\% to produce a
$10^{51} {\rm \, erg}$ explosion for the magnetar model.  For stronger
magnetic dynamos, the angular momentum is too low to produce a disk
around any compact remnant (NS or BH) and the neutron star is not spinning
sufficiently rapidly to produce a GRB.  Clearly, whether or not any
engine works depends upon the angular momentum and it is not clear
that stars can produce GRBs without some method to spin them up.

%
% For one-column wide figures use

Angular momentum is a requirement for all the engines considered in
this paper.  As we shall see in our studies of disk models
(Section~\ref{sec:prog}), different progenitors produce different angular momenta and, hence, will have different observational properties (Sections~\ref{sec:dur},\ref{sec:pred}).  

\section{Progenitors}
\label{sec:prog}

Although the engines behind all of these models are different, the basic progenitors have similar features: a compact remnant with high angular momentum.  In this section, we will review these progenitors, focusing on two broad progenitor classes: ones involving massive stars or ones involving the merger of two compact remnants.

\subsection{Massive Star Progenitors}

Massive star models made the first prediction confirmed by observations: the association of a supernova with long-duration gamma-ray bursts.  For massive star models, the GRB jet is expected to drill through the star and, at the same time, eject the entire star, producing a supernova-like outburst~\citep{macfadyen99}.  No matter how a star is disrupted, whether it be a classic supernova explosion or a jet-driven outburst, an optical display will occur that peaks roughly 10-30\,days after the explosion.  This is simply a consequence of the evolution of the photosphere through the hot ejecta as it expands.  Although additional power sources ($^{56}$Ni decay or magnetar) can augment this emission, the supernova-like
light-curve will occur with or without this additional power source.  With the
observation of supernova 1998bw associated with GRB 980425~\citep{galama99}, massive
stars became the likely candidate for some GRBs and theorists argued 
that many long GRBs should have these supernova-like outbursts.

Less than 0.1\% of all stellar collapses produce gamma-ray bursts~\citep{fryer07,sun15,levan16}.
One aspect of a progenitor model for GRBs is that the system must
normally {\it not} produce a GRB, i.e. it is rare.  Based on our current understanding of supernovae and the initial mass function, roughly 10\% of all massive stars that undergo core-collapse form black holes~\citep{2001ApJ...554..548F}, so even this scenario must only work 1\% of the time.  The explanation for the
rarity of these systems has been that it is difficult to achieve
sufficient angular momentum to form a disk or rapidly-rotating
magnetar.  In the accretion disk models, if there is insufficient 
angular momentum, a disk isn't formed and the engine fails.  The magnetar 
engine predicts a continuous range of drives with different spins.  Superluminous supernovae may be examples of magnetars spinning too slowly to form a GRB~\citep{kasen10,metzger15}.
Another clue to the progenitor may be that, thus far, all
supernovae associated with GRBs appear to be type Ic supernovae\footnote{Type Ic 
are characterized by having no helium lines in their spectra - suggesting little or no 
helium in the ejecta}.  Progenitor models, some favoring NS or BH formation, include:
\begin{itemize}
\item{GRBs only form from the stars with the fastest birth spin rates.
  Fast rotation can cause strong mixing between layers, possibly
  causing the complete burning of the hydrogen and helium
  layers~\citep{yoon05}.  This mixing, and hence homogenization, of
  the star tends to only occur in rapidly rotating stars.  It
  increases the core mass of massive stars, tending to produce black
  holes directly (producing the BHAD engine).}
\item{Spin up by extremely close binaries.  Stars in close binaries
  undergo mass transfer and, in many cases, an expanding star
  envelopes its companion, causing a common envelope phase.  The
  common envelope phase tightens the orbit.  Tidal interactions often 
  spin down the stellar cores but, if the orbit is
  sufficiently tight, tidal locking can spin up the
  stars~\citep{vandenheuvel07}.  With a weak or no dynamo slowing down
  the spin, all 3 (magnetar, NSAD, and BHAD) engines may be produced in 
  this progenitor.}
\item{In a subset of close binaries, the stars undergo a second common
  envelope phase prior to the collapse of the fastest-evolving star
  and, in a subset of these systems, this causes the two helium cores
  to merge, producing a rapidly rotating core~\citep{fryer05}.  With a weak or
  no dynamo slowing down the spin, all engines may work.}
\item{In a subset of systems where the fastest-evolving star collapses
  before its companion expands off the main sequence, the compact
  object can merge to the core of its companion, the perhaps
  poorly-named helium-merger scenario~\citep{fryer98,zhang01}.  High
  accretion rates in this scenario probably bury any magnetic field,
  so this model can only produce disk engines.  The accretion rate
  will bury a magnetar magnetic field, but before the compact remnant
  collapses to form a BH, the NSAD may work.  The BHAD disk will work
  for this system.}
\end{itemize}
Achieving sufficient angular momentum is the driver behind all of
these scenarios.  For massive stars, the evolution of the magnetic 
field in the core depends both on the coupling of different stellar 
layers through, for example, a magnetic dynamo~\citep{heger05}, or 
through mass loss from winds.  Mass loss depends upon metallicity, 
and there is a tendency to argue that angular momentum loss from 
winds is higher at higher metallicities where mass loss from winds is highest, but anisotropies in this mass 
loss make it both difficult to determine the total angular momentum 
loss and the dependence of this angular momentum loss on winds~\citep{georgy13,gagnier19}.
For all but the last scenario, our current
understanding of the magnetic dynamo makes it difficult to get high
enough angular momentum profiles for the magnetar or neutron star
accretion disk models to work.  Indeed, it is difficult for these
models to work even for the black hole accretion disk scenario.  But it 
could be the progenitor is just that rare case where coupling 
does not occur.  Ruling out a scenario is difficult when the event 
rate is as rare as the GRB event rate.  The
helium-merger model is the only current model that definitively
produces sufficient angular momentum.

At this time, none of the models can easily explain the fact that all
of the supernovae associated with GRBs are type Ic~\citep{fryer07}.
Enhanced mixing models have argued that massive stars are more likely
to homogenize~\citep{frey13}.  Black hole accretion disk models require more
massive progenitors (otherwise the star forms a neutron star).  These helium shells of these massive progenitors can undergo strong mixing and burn to carbon and oxygen.  This is, perhaps, an explanation for the predominance 
of type Ic supernovae in GRBs.  This explanation, however, only works 
for the BHAD engine.

Table~\ref{tab:prog} summarizes the possible massive-star progenitors
and their potential engines.  Whether or not the engines work depends
upon the angular momentum in the massive star system.  Magnetic fields
are often invoked as the manner in which different burning layers are
coupled, tying the angular momentum of the core to the outer layers of
the star, slowing down the spin.  In table~\ref{tab:prog}, the
strength of the magnetic dynamo determines this spin (the stronger the
dynamo, the greater the coupling and the slower the spin).

\begin{table}
\caption{Which Progenitors Work For Each Engine: ``no dyn'' $\equiv$ the model works if there is no dynamo (or other viscous force) coupling the burning layers, "weak dyn" $\equiv$ the model works even if there is some coupling between burning layers, ``acc'' refers to the accretion rate.  Note that magnetar engines require that the magnetic field not be buried.}
\label{tab:prog}       % Give a unique label
% For LaTeX tables use
\begin{tabular}{llll}
\hline\noalign{\smallskip}
Progenitor & Magnetar & NSAD & BHAD  \\
\noalign{\smallskip}\hline\noalign{\smallskip}
Rotating Star & no & no & weak dyn \\ 
Tidal Locking & no dyn & no dyn & weak dyn \\
He-He merger & no dyn & no dyn & weak dyn \\
He-merger & no & before BH & yes \\
WD/WD merger & yes & low acc & no \\
WD/NS merger & yes & low acc & BH? \\
WD/BH merger & no & no & low acc \\
NS/NS merger & yes & high acc & high acc \\
NS/BH merger & no & no & high acc \\
\noalign{\smallskip}\hline
\end{tabular}
% Or use
\end{table}

\subsection{Compact Mergers}

Another class of models that produces the conditions for our GRB
engines are the mergers of two compact remnants: BHs,
NSs, or white dwarfs (WDs).  For the magnetar model
and NSAD engines, the merged remnant must be a NS(at least for
some time after the merger) and for the BHAD, the remnant must 
ultimately become a BH.  For the accretion disk models, a disk 
of material must remain behind to power the engine.  These constraints 
rule out mergers of a BH/BH, but we discuss the other mergers here:
\begin{itemize}
\item{WD/WD: Depending on the composition and merger process, the
  merger of two WDs may either ignite into a thermonuclear explosion
  (type Ia Supernova~\citep{livio18}) or collapse to form a neutron
  star (accretion induced collapse - AIC~\citep{fryer99b}).  If it
  collapses, it can produce a dim, but fast supernova with an
  accretion disk around rapidly spinning neutron
  star~\citep{fryer09b,schwab16}.  This progenitor may work for either a
  NSAD or magnetar engine but because it is very unlikely this merger
  will collapse to form a BH and it is not a viable BHAD progenitor.}
\item{WD/(NS or BH): The merger of a white dwarf with either a neutron
  star or black hole will produce a large disk (roughly $10^{10}\,{\rm
    cm}$ in size) around a NS or BH remnant~\citep{fryer99c}.  This merger 
    could spin up a neutron star, producing a magnetar GRB (if the magnetic field is not
  buried or after the magnetic field resurfaces)~\citep{metzger12}.
  The accretion rate will be lower than other mergers for both NSAD
  and BHAD models.}
\item{NS/NS: Neutron star mergers are one of the best-studied GRB
  models.  The merger forms a rapidly spinning neutron star.  The
  neutron star is extended ($\sim 20\, {\rm km}$) after the merger but
  is spinning with $\omega=1500-2000
  Hz$~\citep{ruffert97,rosswog04,rosswog07,fryer15}.  If the engine can
  tap 10-20\% of the rotational energy, it could power a $10^{51} {\rm
    \, erg}$ burst.  For accretion disk systems, disk properties dictate observable features.  The disk is compact with a short lifetime
  (100\,ms), but is expected to be high power (few times $10^{51} {\rm
    \, erg}$)~\citep{popham99}.  The high accretion rates and short
  duration are why theorists classified this as a short GRB
  progenitor~\citep{fryer99a,popham99}.}
\item{NS/BH: NS/BH mergers are similar to NS/NS mergers, but limited to the 
BHAD system.  The power in NS/BH merger depends upon the mass and the
  black hole spin.  The issue is that, especially for the more massive
  black holes, the tidal disruption of the neutron star occurs within
  the innermost stable circular orbit of the black
  hole~\citep{janka99,ruffert00,shibata07}.  Such systems will not
  form a GRB.  Population synthesis calculations suggest that 1\% of
  NS/BH mergers with low-spin black holes ($a<0.6$) form disks and
  40\% of high-spin black holes are able to form
  disks~\citep{belczynski08}.  But these systems can form massive
  accretion disks, producing the most energetic short-duration GRBs.}
\end{itemize}
Table~\ref{tab:prog} summarizes the possible merger progenitors and
their potential engines.  Obviously, which engine works depends upon
the fate of the merged compact remnant: BH versus NS.  For the
magnetar engine, the merger must not lead to rapid accretion that
buries the magnetic field preventing any outburst until the magnetic
field resurfaces.

\section{GRB Durations}
\label{sec:dur}

The durations of our engines can be estimated analytically, providing
constraints on which progenitors (with which engines) can explain the
observed durations.  In this section, we review these estimates.  

For magentar GRBs, the duration can be estimated using pulsar spin-down
models that, in turn are approximate and depend upon assumptions of
the particle acceleration in a pulsar.  Models of increasing
sophistication exist~\citep{contopoulos06}:
\begin{equation}
L_{\rm pulsar} = \frac{B^2 r_{\rm NS}^6 \omega_{\rm NS}^4}{4 c^3}
(\alpha_{\rm psr} sin^2\theta + [1-\omega_{\rm death}/\omega_{\rm NS}]
cos^2\theta)
\end{equation}
where $B$ is the dipole magnetic field, $r_{\rm NS}$ is the neutron
star radius, $\omega_{\rm NS}$ is the neutron star spin, $\alpha_{\rm
  psr}$ is the dipole spin-down efficiency, $\omega_{\rm death}$ is a
function of the magnetic potential gap and the dipole magnetic field
flux, and $\theta$ is the alignment of the spin and dipole magnetic
field axes.  For rough estimates, many just take the leading terms,
assuming $\alpha_{\rm PSR}=1$ and $\theta=90^\circ$:
\begin{equation}
L_{\rm pulsar} = \frac{B^2 r_{\rm NS}^6 \omega_{\rm NS}^4}{4 c^3} \approx 
10^{49} (B/10^{15}G)^2 {\rm erg \, s^{-1}}
\end{equation}
for a 10\,km, millisecond pulsar.  The duration of such an engine is 
approximated by taking the ratio of the rotational energy to the 
pulsar power ($T_{\rm magnetar}=E_{\rm rot}/L_{\rm pulsar}$):
\begin{equation}
T_{\rm magnetar} = 100 (r_{\rm NS}/10\,{\rm km})^{-4} (\omega_{\rm NS})^{-2}
B^{-2} \,s.
\end{equation}
The qualitative properties of magnetar-driven GRBs follows trends in the data: i.e., higher spin magnetars will produce shorter, but more powerful, GRBs, matching the long-soft, short-hard subclasses of bursts.  But this engine struggles to match more detailed data.  For example, it appears that the long bursts typically have more, not less, total energy, arguing that they are produced from faster, not slower, spinning magnetars (contrary to 
the prediction from our simple duration fit).  To make the magnetar model 
fit all GRB populations, we will have to invoke more parameters and features.  
Astrophysicists also do not understand magnetic field generation enough to make strong 
predictions on which progenitors make long versus short bursts or even 
why there is a bimodal distribution of durations.  Simply put, the magnetar model struggles to explain current data on GRB durations and, although the uncertainties 
allow it to fit the current data, the parameters are so poorly known that 
it can not make firm predictions about different properties of short and long bursts.

The disk models are better understood, making firmer predictions.  The
duration of the disk models is set by the lifetime of the disk.  If
the disk is not continuously fed, the duration is the lifetime of the
disk.  If we assume an $\alpha$ disk, the lifetime of the disk
($T_{\rm disk}$) can be estimated by the orbital period of the maximum
extent of the disk divided by $\alpha$:
\begin{equation}
T_{\rm disk}=(2 \pi r^{3/2}_{\rm disk})/(G^{1/2}M_{\rm rem}^{1/2}\alpha)
\end{equation}
where $r_{\rm disk}$ is the radius of the disk, $G$ is the gravitational 
constant, and $M_{\rm rem}$ is the remnant mass.  For a disk formed 
from material with specific angular momentum $j$, the timescale for the 
disk accretion is:
\begin{eqnarray}
T_{\rm disk} & = & (2 \pi j^{3})/(G^2 M_{\rm rem}^2 \alpha) \\
 & = & 4 (j/10^{17} {\rm cm^2 \, s^{-1}})^3 (3 M_\odot/M_{\rm rem})^2
(0.01/\alpha) \, s. \nonumber
\end{eqnarray}
For a typical neutron star merger, the disk lifetime is roughly
100\.ms, fitting the average timescale of short-duration bursts.  For
helium-merger models, the specific angular momentum can be $10^{18} \,
{\rm cm^2 s^{-1}}$, producing very long (more than 1000\,s) GRBs.  
With the discovery of weak, ultra-long bursts, interest in the He-merger 
progenitor has increased.

This disk accretion timescale is appropriate for compact mergers where
there is no way to feed the disk.  But for massive stars, the disk is
replenished by further accretion onto the disk from the star which
continues either until the star is disrupted or it completely accretes
onto the compact remnant.  The free-fall time ($T_{\rm free-fall}$) of
the star onto the compact remnant places an upper limit on the
duration of the engine:
\begin{eqnarray}
T_{\rm free-fall} & = & \pi r_{\rm star}^{3/2}/(8 G M_{\rm star})^{1/2}
\\ & \approx & 35\,s (r_{\rm star}/10^{10} {\rm \, cm})^{3/2}
(8\,M_\odot/M_{\rm star})^{1/2} \nonumber
\end{eqnarray}
where $r_{\rm star}$ is the radius of the stellar He or CO core and
$M_{\rm star}$ is the mass of this core.  Depending upon the
compactness of the core, the timescale for CO core collapse can 
be as high as a few hundred seconds.  For longer-lived GRBs, the 
star must include part of the helium layer, a problem if these 
ultra-long bursts also only have associated type Ic, instead of 
Ib, supernovae.

\cite{popham99} reviewed different progenitors, their durations and
their power (energy deposited per unit time - ${\rm erg \, s^{-1}}$).  Since this time, better models for the disks have been
produced and we can use these models and the \cite{popham99} analysis
to estimate powers and durations of the different progenitors.  Table~\ref{tab:dur} shows the power and durations for the progenitors in this study combining 
analytic prescriptions, disk models and our current understanding of the 
progenitor evolution.  The prediction is the same for the massive star models at this time, because the models are not yet sufficiently detailed to differentiate the 
progenitors.  For massive stars, the lower time limit is set by the 
time the jet requires to punch through the star ($\approx R_{\rm star}/c \sim 1 {\rm s}$ for C/O cores).  The upper limit is 
set by the infall time of a C/O core or, in the case of NSAD models, the 
collapse time of the neutron star.

\begin{table}
\caption{Durations and Power for Accretion Disk Models}
\label{tab:dur}       % Give a unique label
% For LaTeX tables use
\begin{tabular}{lll}
\hline\noalign{\smallskip}
Progenitor & Power & Duration  \\
\noalign{\smallskip}\hline\noalign{\smallskip}
{\bf BHAD Models} & & \\
Rotating Star & $\sim 10^{50} {\rm \, erg \, s^{-1}}$ & $1-300\,s$ \\
Tidal Locking & $\sim 10^{50} {\rm \, erg \, s^{-1}}$ & $1-300\,s$ \\
He-He merger & $\sim 10^{50} {\rm \, erg \, s^{-1}}$ & $1-300\,s$ \\
He-merger & $\sim 10^{50} {\rm \, erg \, s^{-1}}$  & $300-10^{4}\,s$ \\
WD/BH merger & $\sim 10^{49} {\rm \, erg \, s^{-1}}$  & $300-10^{4}\,s$ \\
NS/NS merger & $\sim 10^{51} {\rm \, erg \, s^{-1}}$  & $<1\,s$ \\
NS/BH merger & $\sim 10^{51} {\rm \, erg \, s^{-1}}$  & $<1\,s$ \\
\noalign{\smallskip}\hline
{\bf NSAD Models} & & \\
Rotating Star & $\sim 10^{50} {\rm \, erg \, s^{-1}}$ & $1-10\,s$ \\
Tidal Locking & $\sim 10^{50} {\rm \, erg \, s^{-1}}$ & $1-10\,s$ \\
He-He merger & $\sim 10^{50} {\rm \, erg \, s^{-1}}$ & $1-10\,s$ \\
He-merger & $\sim 10^{50} {\rm \, erg \, s^{-1}}$  & $300-10^{4}\,s$ \\
WD/WD merger & $\sim 10^{49} {\rm \, erg \, s^{-1}}$  & $300-10^{4}\,s$ \\
WD/NS merger & $\sim 10^{49} {\rm \, erg \, s^{-1}}$  & $300-10^{4}\,s$ \\
NS/NS merger & $\sim 10^{51} {\rm \, erg \, s^{-1}}$  & $<1\,s$ \\
\noalign{\smallskip}\hline
\end{tabular}
% Or use
% \vspace*{5cm}  % with the correct table height
\end{table}

\section{Progenitor/Engine Predictions}
\label{sec:pred}

Aside from the association between long duration GRBs and supernovae,
most of the constraints we have discussed thusfar are
``post-dictions''; that is, theorists ensuring that their models fit
existing data.  The different progenitor and engine models have made a
series of predictions.  For some engines, the subsequent observations 
begin to rule out some engines.  For others, the observations have 
been confirmed, strengthening the case for these models/engines.

\subsection{Magnetar Engines and Late-Time Emission}

Because we know very little about the neutron star spin and magnetic
fields, it is difficult to tie progenitors to predictions about the
magnetar engine.  As we mentioned in the discussion on durations, we 
do not understand the generation of magnetic fields well enough 
to tie strong magnetic fields to particular progenitors.  But 
the magnetar model does make predictions based on the engine alone.  
If a magnetar is driving the GRB, there should be late-time emission in the afterglow (e.g. radio) (unless the star collapses to a black hole). This emission is not seen in short bursts~\citep{fong16}, arguing 
against the magnetar engine for these bursts.  In addition, observations from
GW170817 also have shown that no magnetar existed at late times although it is possible the magnetar collapsed to a black hole after driving a jet or that a normal field ($10^{12} {\rm Gauss}$) neutron star exists in this merger~\citep{piro19}.  It is clear that the mangetar engine must overcome some challenges to explain the current set of short-duration GRB observations.

\subsection{Distributions of Disk Engines}

Because we can tie properties of the accretion disk models to specific
progenitors, we can use our understanding of these progenitors to make
predictions on further properties of these models.  As we discussed in
section~\ref{sec:dur}, short-duration GRBs are expected to arise from
NS/NS and NS/BH mergers.  Long bursts are expected to arise from
massive star models, He-mergers or WD/(BH or NS) mergers.  But the
bulk of long bursts are expected to arise from massive stars.  Within
this framework, scientists began to make predictions for different
bursts: long bursts should form in star-forming galaxies whereas short
bursts should occur in all galaxy types.

Theory made further predictions based on the accretion disk paradigm.
If short bursts are only produced by NS/NS and NS/BH mergers, the
kicks imparted onto NSs at formation cause these binaries to be
ejected from their star forming regions (and, in some cases, even
their host galaxies).  Theorists made predictions for the distribution
of offsets of these mergers~\citep{fryer99a,bloom99}, arguing
that short duration GRBs should be much more spatially extended than
supernovae.  If the BHAD engine is correct, long-duration bursts
should form from the most massive stars, located close to young star
forming regions.

A growing set of observations seem to support this paradigm: long bursts appear to be concentrated near the peak emission in star-forming galaxies, are even more centrally located than supernovae~\citep{vreeswijk01,bloom03,fruchter06}.
Full confirmation of the accretion disk paradigm required observations demonstrating short GRBs arise from mergers, e.g. a large offset distribution.  The Swift satellite pinpointed the
position of short GRBs sufficiently to allow good follow-up
observations of these bursts, proving that short-duration bursts were
not only more distributed than their counterparts, but some were found
that are outside their host galaxy~\citep{fong13a,fong13b}.  

GW170817 cemented the neutron star merger scenario for short bursts, but the
observations were more detailed than is often realized.  The
gravitational wave detection~\citep{abbott17} proved that a compact
merger (from the inferred masses, a NS/NS merger) event occured.  The
corresponding gamma-ray observations~\citep{savchenko17} showed that
gamma-rays did arise from this merger.  X-ray afterglow
observations, e.g.~\citep{troja17}, suggested an off-axis GRB and the
radio argued strongly for the presence of a relativistic
jet~\citep{mooley18}.  All the evidence together argues strongly that 
GW170817 was a true short GRB from a neutron star merger.  

By assuming the accretion disk paradigm is behind GRBs, astrophysicists can predict which progenitors are behind different types of bursts:  massive stars dominated the progenitors of long bursts and NS/NS and NS/BH mergers produce short bursts.  The prediction of this paradigm allowed astronomers to use population characteristics of the different burst types to either confirm or rule out this engine.  The subsequent 20 years of observations, including the recent GW170817 observations, have supported the predictions of this engine.  With such a strong confirmation, the accretion disk paradigm remains the leading GRB class of engines.

\subsection{Mergers and the r-Process}
\label{sec:rproc}

During the merger, matter is ejected both during the tidal disruption of the neutron star(s) and through winds during the subsequent disk accretion onto the core~\citep{metzger12}.  The neutron-rich dynamical ejecta from BH/NS and NS/NS mergers were suggested as possible r-process sites~\citep{lattimer74}.  In addition 
to playing an important role in the synthesis of heavy elements in the universe, this ejecta produces a fast transient.  Like a thermonuclear supernova, the decay of the neutron-rich, radioactive isotopes produced in these mergers can power a transient from which we can probe the ejecta properties.  Several 
potential infra-red detections were made prior to GW170817~\citep{berger13,tanvir13}, but many of these detections had corresponding X-ray flares, suggesting a shock interaction instead 
of a true kilonova observation~\citep{kasliwal17}.  The optical 
and infra-red observations of GW170817 provided the first definitive 
observation of the ejecta from neutron star mergers.

Before we discuss the results of the observations, let us review what
was known from theory.  Simulations demonstrated that the tidal ejecta
were sufficiently neutron rich to produce heavy
r-process~\citep{freiburghaus99,goriely11,2011ApJ...736L..21R,korobkin12,Richers15,2016ApJ...830...12M,sekiguchi16,radice16,shibata17,bovard17,foucart18}.
Scientists found that the total amount of neutron-rich ejecta depends
upon the relative neutron star masses of the two compact
remnants~\citep{korobkin12}.  However, uncertainties in the
simulations (e.g. implementation of the physics and numerical
artifacts) are fairly large and the results from current calculations 
range from $0.001-0.05\,M_\odot$.  NS/BH mergers have a much wider 
range of dynamical ejecta masses ranging from no ejecta whatsoever when the 
the black hole is large and the neutron star is tidally disrupted within 
the innermost stable circular orbit up to $\sim 0.1\,M_\odot$ for smaller-mass black holes.

Some basic trends exist with the disk ejection.  If the remnant
remains a neutron star, ``wind'' ejecta includes both disk winds and
accretion outflows (on par with the re-ejection of mass in supernova
fallback~\citep{fryer09})~\citep{metzger12}.  If the remnant is a BH, this
latter outflow does not occur and we expect more ejecta from systems with 
NS cores than those with BH cores.  Neutrinos can reset the neutron fraction in the ejecta to equal numbers of neutrons and protons and it is believed that systems with NS cores (and their resultant neutrino flux) will be reset further than those with BH cores.  
But just as with the dynamical ejecta, physic implementation 
and numerical artifacts dominate the solutions, and the wind ejecta 
masses range from $0.001-0.05\,M_\odot$ with electron fractions [$Y_e = n_e/(n_e+n_n)$ where $n_e, n_n$ are the numbers of electrons and neutrons respectively] ranging from slightly above dynamical ejecta (0.1-0.25to 0.35-0.4.  The electron fraction plays a major role in 
dictating the yields\footnote{The yields of any explosive event depend upon the electron fraction, the entropy and the density/temperature evolution~\citep[see, for example][]{2015ApJ...815...82L,2010ApJS..191...66M}.  The density/temperature profiles are generally well fit by analytic profiles.  For the variations of these parameters in neutron star mergers, these latter two parameters do not have as strong an effect as the electron/neutron fraction on the broad r-process yields.  However, all three must be understood to produce exact yields.}.  For high electron fractions, the ejecta does not produce heavy r-process elements, and the ejecta can be dominated by iron peak elements.  At this time, simulation differences are larger than the trends.

\begin{figure}
% Use the relevant command for your figure-insertion program
% to insert the figure file.
% For example, with the option graphics use
\resizebox{0.48\textwidth}{!}{%
  \includegraphics{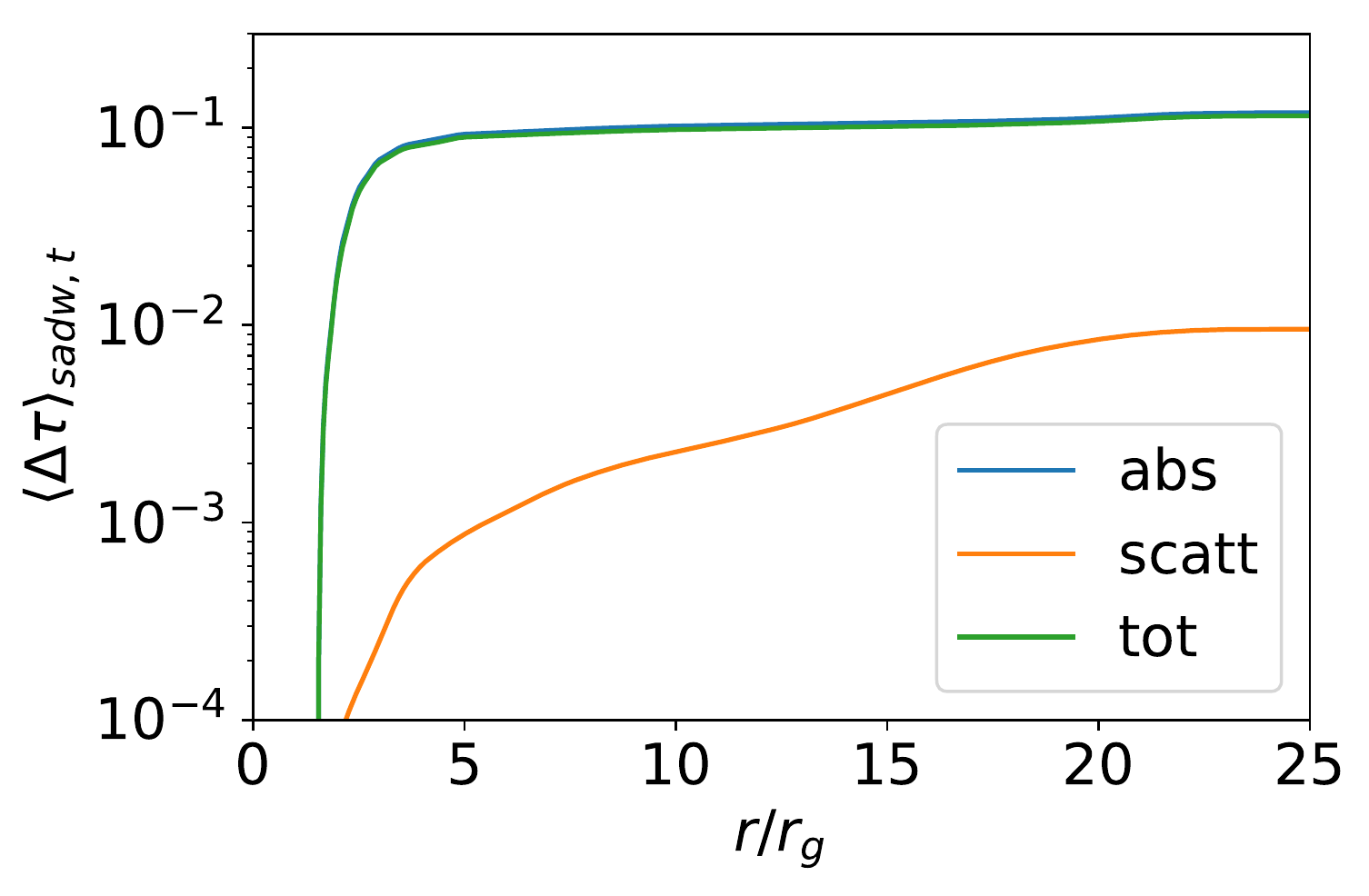}
}
\caption{Average absorption, scattering, and total optical depth that a neutrino emitted in the inner region of a post-neutron-star-merger disk passes through as a function of distance from the black hole in gravitational radii. The optical depth is averaged both over spherical shells at each radius and over time, from 20ms to 27ms in a two-dimensional simulation.}
\label{fig:disktau} 
\end{figure}

Calculating the electron fraction, and hence final yields, from the late-time wind ejecta depends upon many modeling and physics uncertainties including neutrino transport and its coupling to the hydrodynamics, magnetic fields, dense nuclear equations of state and neutrino physics.  Although the number of models of this wind ejecta continues to increase~\citep{FoucartPostMerger,NouriPostMerger,SiegelMetzger3DBNS,FernandezLongTermGRMHD,nubhlight}, these uncertainties still produce ejecta with a wide range of electron fraction.  To understand the extent of these uncertainties, we study just one aspect, the neutrino transport.  Figure \ref{fig:disktau} shows the average absorption, scattering, and total optical depths that a neutrino emitted in the inner region of a post-neutron-star-merger disk passes through as a function of distance from the central black hole from a disk model using the $\nu\texttt{bhlight}$ code~\citep{nubhlight}. A vanishing optical depth would imply that optically thin cooling is appropriate, whereas an infinite optical depth would imply neutrinos never escape. Although the disk is optically thin, the optical depth is not negligible. Free-streaming will not capture all of the physics.

%
% For one-column wide figures use

\begin{figure}
% Use the relevant command for your figure-insertion program
% to insert the figure file.
% For example, with the option graphics use
\resizebox{0.48\textwidth}{!}{%
  \includegraphics{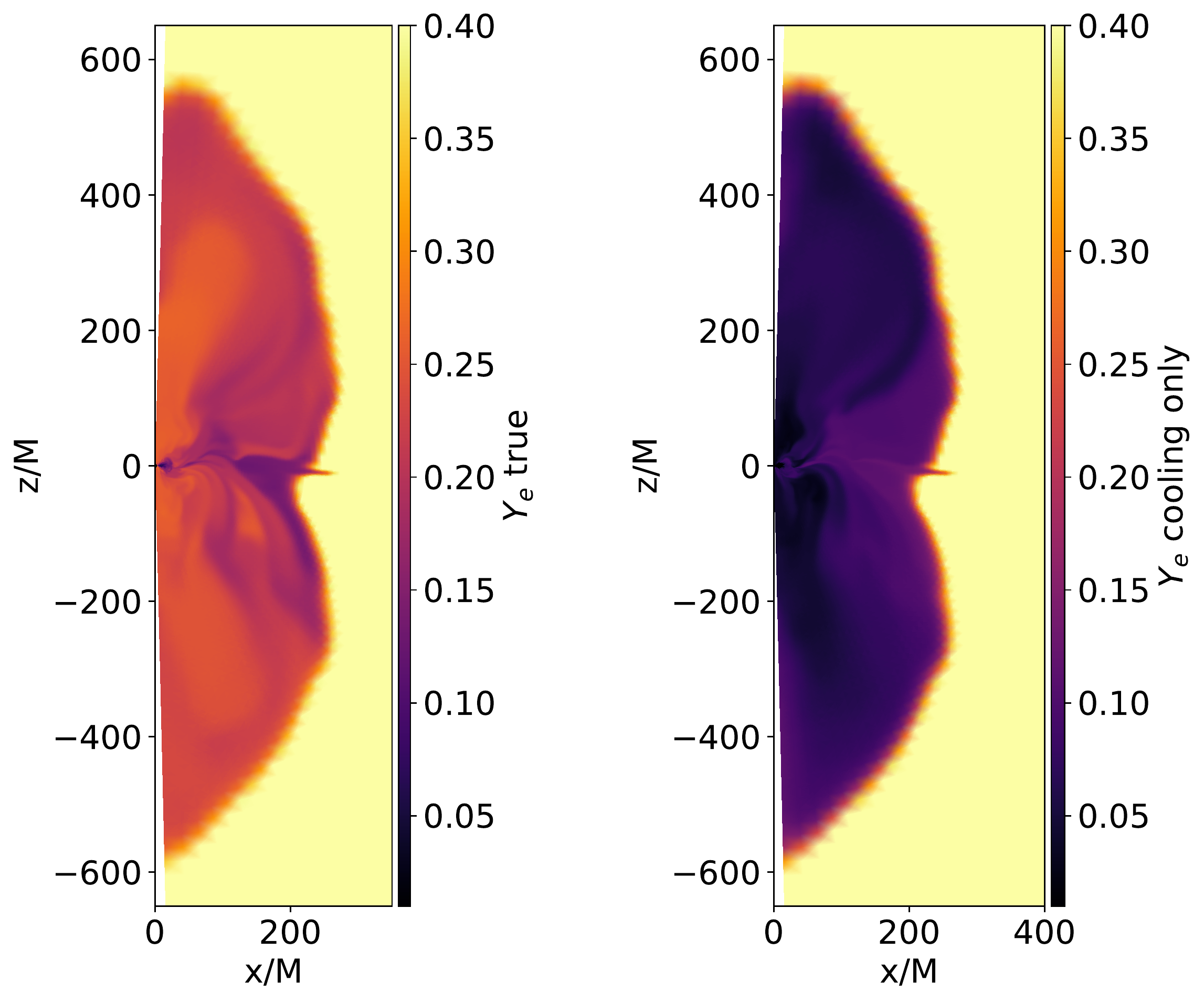}
}
\caption{Electron fraction in the outflow after 27ms from a two-dimensional simulation of a post-neutron-star-merger disk. Left: electron fraction if one solves the Boltzmann equation for the neutrinos. Right: electron fraction if one assumes optically thin cooling and no absorption.}
\label{fig:diskye} 
\end{figure}

Figure \ref{fig:diskye} shows the electron fraction of the ejecta from a black hole accretion disk using the $\nu\texttt{bhlight}$ code, both with neutrino transport (using Monte Carlo methods) and a free streaming prescription. The left panel shows the electron fraction using full transport, accounting for absorption, emission, and scattering. With full transport, the electron fraction varies dramatically in space, reaching relatively large values. The right panel shows the electron fraction if one instead assumes optically thin cooling, with no absorption or scattering. With only cooling, the electron fraction is universally low and the outflow is extremely neutron rich. These two outflows may produce dramatically different nucleosynthetic yields.  These will produce different light-curves and contribute differently to the r-process.

%
% For one-column wide figures use

\begin{figure}
% Use the relevant command for your figure-insertion program
% to insert the figure file.
% For example, with the option graphics use
\resizebox{0.48\textwidth}{!}{%
  \includegraphics{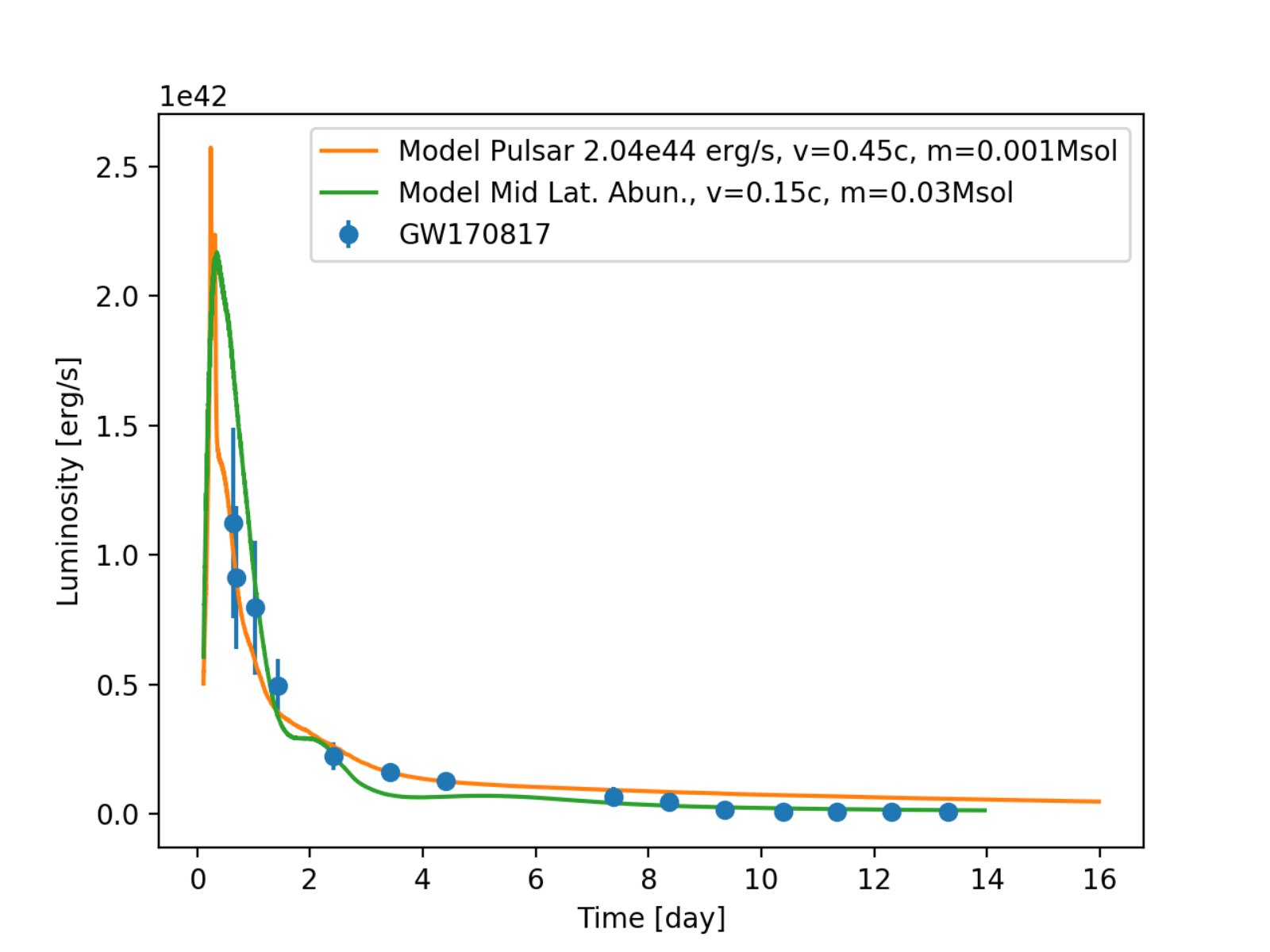}
}
\caption{Bolometric luminosity versus for two different kilonova
  models: the W2 wind model~\citep{wollaeger18} with 0.03\,M$_\odot$ of
  wind ejecta (high electron fraction) moving at 0.15 the speed of
  light, and a wind model with 0.001\,M$_\odot$}
\label{fig:lpuls} 
\end{figure}

The combination of both neutron rich dynamical ejecta (producing 
the r-process) and neutron poor wind ejecta led theorists to predict 
two components in the light curve \citep{metzger14}.  Ultraviolet, optical 
and infra-red observations of this ejecta can be used to determine 
the relative fractions neutron-rich material.  Unfortunately, GW170817 has demonstrated just how difficult it is to produce firm ejecta masses from the light-curves and the range of ejecta masses fit to the observations are nearly as large as the uncertainties in the models~\citep{cote18}\footnote{Perhaps not surprising given the difficulty in estimating supernova ejecta masses from light-curves and spectra.}.  In addition to the ejecta mass, the light-curve depends upon the opacities and their coupling to transport, understanding of aspects of the ejecta (morphology, velocity and density distributions, exact composition, ...), and other energy sources.  For example, although, as mentioned above, we can show that GW170817 does not have a long-lived magnetar, observations do not rule out a long-lived pulsar~\citep{piro19}.  Such a pulsar will affect the early time
emission.  Figure~\ref{fig:lpuls} shows the bolometric light curve of
GW170817 matched by two different models: a massive wind (high
electron fraction) ejecta ($0.03$\,M$_\odot$) simulation and a
low-mass wind ejecta ($0.001$\,M$_\odot$) with a $10^{12} {\rm \, G}$
pulsar simulation.  Both fit the observations, but the difference in 
ejecta mass is a factor of 30.  Without detailed modeling (hopefully 
identifying discriminating observables), what we can learn from 
even multi-diagnostic signals is limited.  But some general trends exist:  if the remnant remains a neutron star, the neutrinos from this neutrino star will raise the electron fraction of the ejecta, tending to produce more iron peak and 1st r-process elements.  This produces a stronger optical/UV signal, especially in the first day or two after the merger.  If the core collapses to a black hole, the electron fraction is lower producing heavier r-process elements whose opacities make a redder transient.

%
% For one-column wide figures use

Whether or not mergers can be the dominant source of the r-process depends both on the rate of these mergers and the amount of neutron-rich ejecta~\citep{cote17}.  Prior to GW170817, the strongest constraints on the rate of compact mergers came from observations of close binary pulsar systems.  Pulsar binaries like the Hulse-Taylor pulsar
could be used to estimate the rate of mergers~\citep{kalogera04}.  But selection effects (beaming, recycling efficiencies, kicks) made it difficult to place strong constraints on the rate.  Assuming mergers produce 
short GRBs, astrophysicists could also predict a merger rate but, again, 
biases (uncertain beaming) made it extremely difficult to determine 
an accurate rate~\citep{chen13}.  Theory is equally uncertain.
Errors in the rate that span many orders of magnitude,
allowing the potential for mergers to either dominate the r-process
production in the universe to scenarios where mergers make up only a
few percent of the r-process in the universe~\citep{fryer99a,dominik12}.  The observational uncertainties from gravitational wave detections are much smaller and, even though GW170817 is only one event, it is already placing among the 
most stringent constraints on neutron star merger rates and the role of mergers in producing the heavy elements in the universe~\citep{cote19}.

\section{For the Future}
\label{sec:future}

There is no doubt that GW170817 dramatically improved our understanding of short-duration GRBs.  Neutron star merger rate estimates based on GW170817 are the most stringent to date.  Coupled with optical and infra-red observations, GW170817 provided the first definitive evidence of ejecta from the merger.  
Coupled with gamma-ray, X-ray and radio observations, GW170817 
places some of the strongest evidence that mergers produce 
GRBs.  In all cases, GW170817 is adding to the 50-year set 
of data that has led scientists to the current standard models 
behind GRBs.  We conclude with a description of what the future 
holds for GRB observations.

\subsection{Understanding GRB Properties}

Gravitational waves have already made a major contribution to our 
understanding of short duration gamma-ray bursts.
Gravitational wave detections alone will be able to pinpoint the NS merger rate over the next few years.  The biggest uncertainty in the rate estimate from GW170817 is the fact that, at this time, we have only one event.  However, as aLIGO becomes increasingly sensitive, the rate of merger detections will increase (although it may get more difficult to obtain data across the electromagnetic spectrum) and the sample of NS/NS and even NS/BH should grow dramatically 
with time.  aLIGO alone will tightly constrain the rate of mergers.

Multi-messenger astronomy enhances the science we can learn.  As we build up our observational database of merger detections with concurrent gamma-ray measurements, we will improve our understanding of gamma-ray bursts.  For example, if we can pin down mergers as the primary source of short-duration GRBs, we can use the rate of mergers to constrain the beaming of these outbursts.  A broader set of diagnostics will help us study the jet in
more detail, i.e. the gravitational wave signal and the kilonova signal place limits on the viewing angle (the angle between the orbital angular momentum access and our line of site) of the burst.  GW170817 provided a 
first probe of the off-axis structure of the jet
producing the gamma-ray burst.  As we build up our sample of
well-studied mergers with a range of viewing angles, we can probe in
detail the GRB jet structure.

Coupled with UVOIR measurements, LIGO detections will also grow the sample of ejecta detections.  As we discussed in section~\ref{sec:rproc}, the UV, optical and infra-red measurements are produced by the ejecta in neutron star mergers and observations of this kilonova have been used to determine the characteristics of this ejecta.  Unfortunately, uncertainties in this modeling make it difficult, with one event, to pinpoint the exact yields and, currently, a range of results exist for the broad set of analyses~\citep{cote18}.  However, as the number of events increase and the models improve, scientists will be able to break the current degeneracies, obtaining more exact yields.
This information can be applied to events beyond the LIGO detection threshold.  For example, scientists have taken the properties optical and infra-red emission (arising from decay energy in the ejecta) to see if any other GRBs have similar properties~\citep{rossi19}.  As we detect more 
nearby events (with joint gravitational wave and electromagnetic wave signals), we can determine the range of emission arising from these mergers.  Coupled with detailed models, these observations can probe the characteristics (velocities, density distribution, and 
composition) of the ejecta.  These studies will determine the role mergers play in producing r-process elements in the universe and provide a much more detailed picture of r-process production.  With detailed 
ejecta and light-curve models, astrophysicists will be able to differentiate NSAD and BHAD engines.  And, as we shall discuss below, by determining the fate of the core, we probe the behavior of matter at nuclear densities.

Bear in mind these studies rely on a basic understanding of the
physics behind these outbursts and modelers must deal with both
physics and numerical uncertainties.  For example, the exact nature of
the nuclear physics can alter the yield, changing the energy
deposition and the light-curve (for the same ejecta mass)~\citep{zhu18}.
Understanding and minimizing the nuclear physics uncertainties that
affect the optical and infra-red light-curves are essential to using
observations to constrain ejecta properties.  Other physics effects
are also important.  In section~\ref{sec:pred}, we already reviewed many 
of the uncertainties:  additional energy sources, opacities and opacity implementations into transport, ejecta properties, neutrino and nuclear physics.  Considerable multi-physics work is necessary to fully take advantage of the upcoming data.

Gravitational waves, by themselves, can be used to probe the neutron
star equation of state.  But as the number of well-studied (with
gravitational wave and broad electromagnetic coverage) events
increase, we can further probe the properties of this equation of
state~\citep{fryer15}.  These probes rely upon detailed models studying
the regularization of the angular momentum (it is believed that the
gradients in spin will disappear in the core) and the neutrino cooling
will quickly reduce the thermal pressure.  Comparing different electromagnetic 
signals from these mergers, coupled with gravitational wave constraints on 
the NS component masses, can probe the maximum neutron star or black hole mass.

Gravitational waves have the potential to place strong constraints on the engines and physics behind long-duration gamma-ray bursts.  But, for most of the long-duration GRB progenitors, the gravitational wave signal is likely to be much weaker than neutron star merger progenitors.  For massive star progenitors, signal is likely to be similar to those of supernovae which, for the most part, produce signals that are only detectable for events occuring in the Milky Way~\citep{2011LRR....14....1F}.  For the engine to work, the rotation of these systems must be high and this can lead to higher gravitational wave signals that may, with next generation detections, be detectable out to the Virgo cluster~\citep{2002ApJ...565..430F,2011LRR....14....1F}.  For progenitors invoking mergers with helium stars or white dwarfs produce low-frequency signals in the LISA band, but will also only be detectable in the Milky Way.  Given the rarity of these events, it is unlikely that gravitational-wave detectors will detect a long-duration burst in the near future.  However, if such a detection would occur, gravitational waves would be an ideal probe of the details of the progenitor:  progenitor scenario and its characteristics.

\subsection{Predictions with Redshift}

By increasing our understanding of the properties of GRBs, we can begin to use these powerful explosions to probe the early universe.  But first, we must understand the evolution of these bursts with redshift.  One of the primary differences in the evolution of progenitors with redshift is the difference in the amount of metals.  One of the exciting prospects of well-studied events in the nearby universe is to use them to study metallicity effects so that we may make predictions for these high redshift bursts.

Metals are important for cooling gas in star formation.  Some cosmological models argued that stars without this cooling would not form binaries, arguing population III stars could not form the binary progenitors needed to produce GRBs, decreasing the fraction GRBs at high redshift~\citep{belczynski10}.  However, once numerical issues with the ENZO code were fixed~\citep{passy12}, their population III stars began forming binaries.  Even so, it is still believed that lower cooling will lead to more massive stars, on average, possibly flattening the initial mass function which would increase the fraction of GRBs at high redshift~\citep{lloydronning02,2016A&A...587A..40P}.  Absorption from metals can also lead to larger giant envelopes and increased mass loss, especially for stars with hydrogen envelopes.  
For long bursts, \cite{young07} used these differences to predict the distribution 
of GRBs as a function of redshift for different progenitors.  A similar study 
for short bursts was done by \cite{dominik13}.

The past studies folded prescriptions for metallicity evolution as
well as prescriptions for stars and star formation, typically focusing
on rates.  Well-studied nearby systems (that include both gravitational 
wave and broad spectral data) can be used to probe the variation in 
GRB properties as a function of metallicity.  In this section, we 
will discuss the current predictions for this evolution.  
For each progenitor, \cite{fryer99a} studied the rate dependence on 
the initial mass function, the stellar radius, and mass loss.  From
this study, we can calculate a broad set of GRB properties as a
function of redshift: duration, power, circumstellar properties and 
the rate.  Especially for merger events, timing (tighter mergers merge 
more quickly) can also alter the results and we will include these differences 
as well.

\begin{figure}
% Use the relevant command for your figure-insertion program
% to insert the figure file.
% For example, with the option graphics use
\resizebox{0.48\textwidth}{!}{%
  \includegraphics{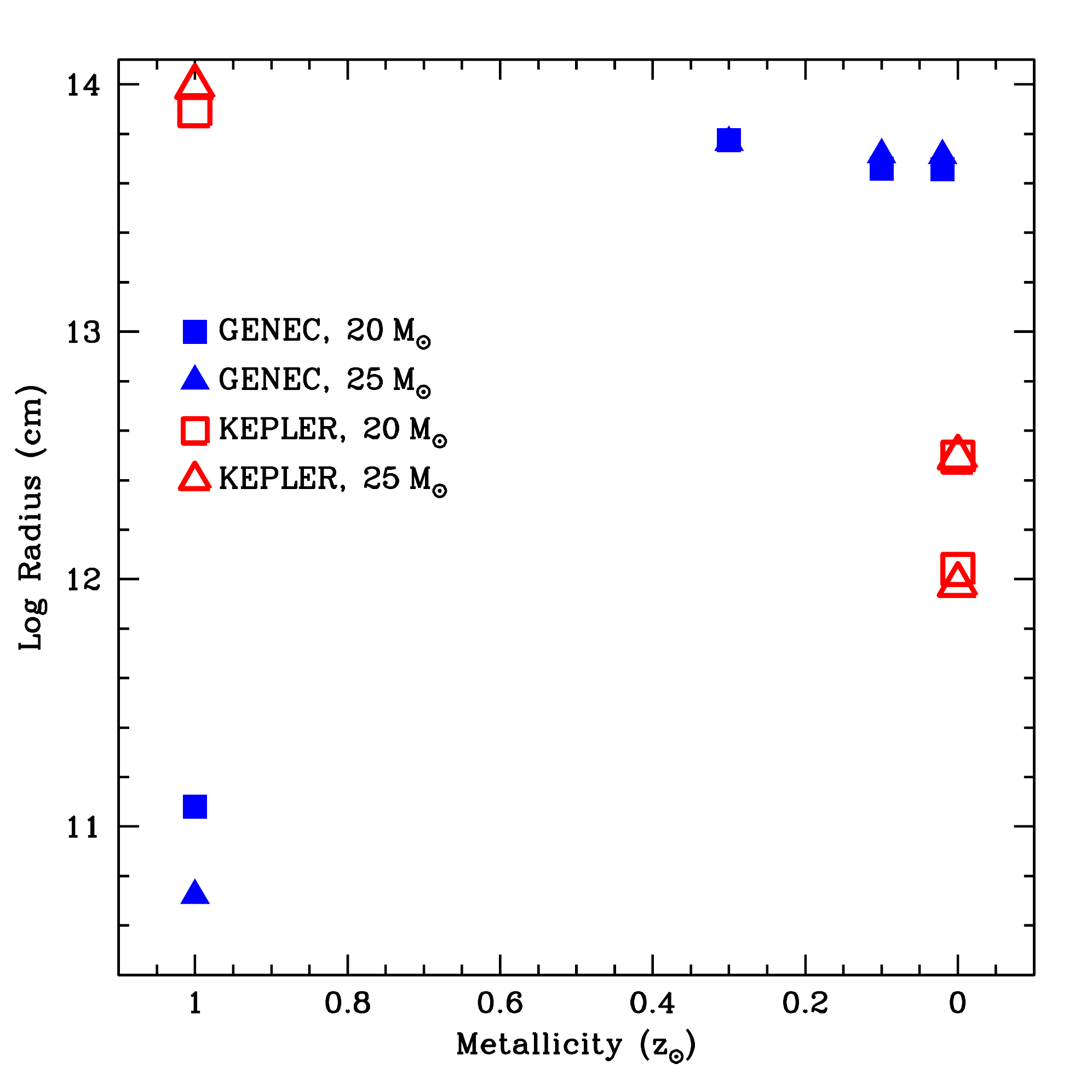}
}
\caption{Stellar radius at collapse for 20 and 25\,M$_\odot$ zero-age main sequence 
stars as a function of metallicity for both GENEC~\citep{belczynski18} and KEPLER~\citep{woosley02} codes.  The GENEC stars lose their entire hydrogen envelopes 
at solar metallicity, making compact helium stars.  For all other models, the radii 
decrease with metallicity.  The most dramatic effect occurs below roughly 1/100th 
solar metallicity.  Population III stars will have radii that are more than an 
order of magnitude lower than stars at 1/10th solar metallicity.}
\label{fig:radii} 
\end{figure}

Before we review individual progenitors and their properties, let us review 
how the mass function, mass loss, and stellar radii effect stellar and binary 
evolution:
\begin{itemize}
\item{{\bf Initial Mass Function:}  Cooling from metals allows smaller proto-stellar 
cores to collapse to form stars and the tendency is for these cores to produce 
lower-mass stars.  Especially for Pop III stars, where this cooling is absent, 
we expect an initial mass function that is highly skewed toward massive stars.   
By itself, this suggests that a much larger fraction of stars will form GRBs 
at high redshift~\citep{lloydronning02}.  Across all progenitor models (except 
for WD/WD mergers), the rate increases with a flattening of the IMF, but 
the rates of BH systems (black hole collapsars, BH/(NS,WD) mergers) increase 
more rapidly than NS systems~\citep{fryer99a}.}
\item{{\bf Stellar Radius:} The amount of expansion in the giant phase
  of a massive star depends upon the opacity.  Since metals dominate
  this opacity, higher metallicity stars will have larger radii.
  Figure~\ref{fig:radii} shows the radii (at collapse - this is not necessarily 
  the maximum radius, but it is generally close to it) for 20 and 25\,M$_\odot$ 
  stars using the GENEC~\citep{GENEC1,GENEC2,GENEC3} and KEPLER~\citep{woosley02} as 
  a function of metallicity.  For the solar metallicity GENEC models, the 
  radius is small because winds eject all of the hydrogen envelope.  For 
  models that retain their hydrogen envelope (all KEPLER models, models below 
  solar metallicity with the GENEC code), the radius decreases with metallicity.  This 
  effect is most dramatic below roughly 1/100th solar metallicity.  Many of the
  binary progenitors for GRBs require common envelope interactions to
  tighten the orbit.  If the initial orbital separation is the same,
  decreasing the radius of the giant star will decrease the number of
  systems that undergo common envelop evolution, lowering the fraction
  of progenitors.  However, this effect is muted by the fact that the
  widest separation systems undergoing common envelope evolution tend
  not to form GRB progenitors: e.g. for compact object progenitors, these wider
  binaries tend to get disrupted during compact object formation and/or
  remain wide and do not merge within a Hubble time.}
\item{{\bf Mass Loss:} Stellar winds are typically believed to be driven 
by atomic lines and higher metallicity stars have stronger winds.  If the 
mass loss is sufficiently extensive to alter the mass of the stellar core 
at collapse, it can change its fate:  e.g. the collapse can form a NS instead 
of a black hole.  This is important for metallicities above 1/10th solar, but, below 
this metallicity, the mass loss from winds is minimal and does not play a major role in altering the fate of the massive star.  But as the metallicity increases from 1/100th solar up to solar, the relative fraction of BH systems decreases and the fraction of NS systems increases.}
\end{itemize} 

%
% For one-column wide figures use

%
% For one-column wide figures use
\begin{figure}
% Use the relevant command for your figure-insertion program
% to insert the figure file.
% For example, with the option graphics use
\resizebox{0.48\textwidth}{!}{%
  \includegraphics{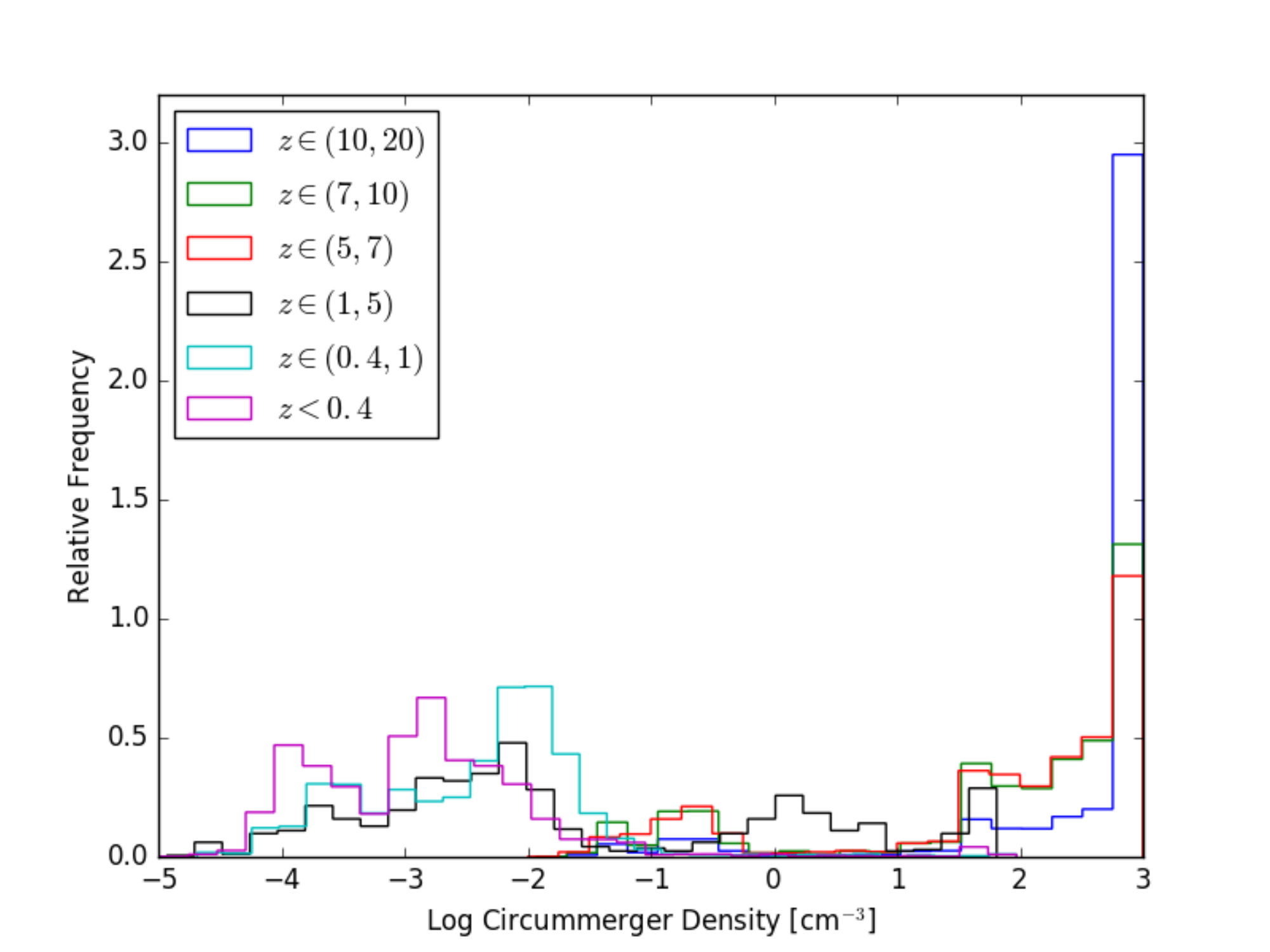}
}
\caption{Distribution of CSM densities as a function of redshift band for double neutron star mergers.  Below a redshift of 1, there is a bimodal peak of the CSM density for these mergers.  The low densities occur for binary with long merger times where the systematic velocity can carry the binary well away from its formation site (and even beyond its host galaxy).  The high densities occur for binaries with short merger times.  The densities become increasingly higher at high redshift because the only systems that have merged at this time have short merger times and hence the systems are found preferentially in their high-density star forming regions.}
\label{fig:nsmerg} 
\end{figure}

\begin{figure}
% Use the relevant command for your figure-insertion program
% to insert the figure file.
% For example, with the option graphics use
  \includegraphics[width=3.4in,height=3.5in]{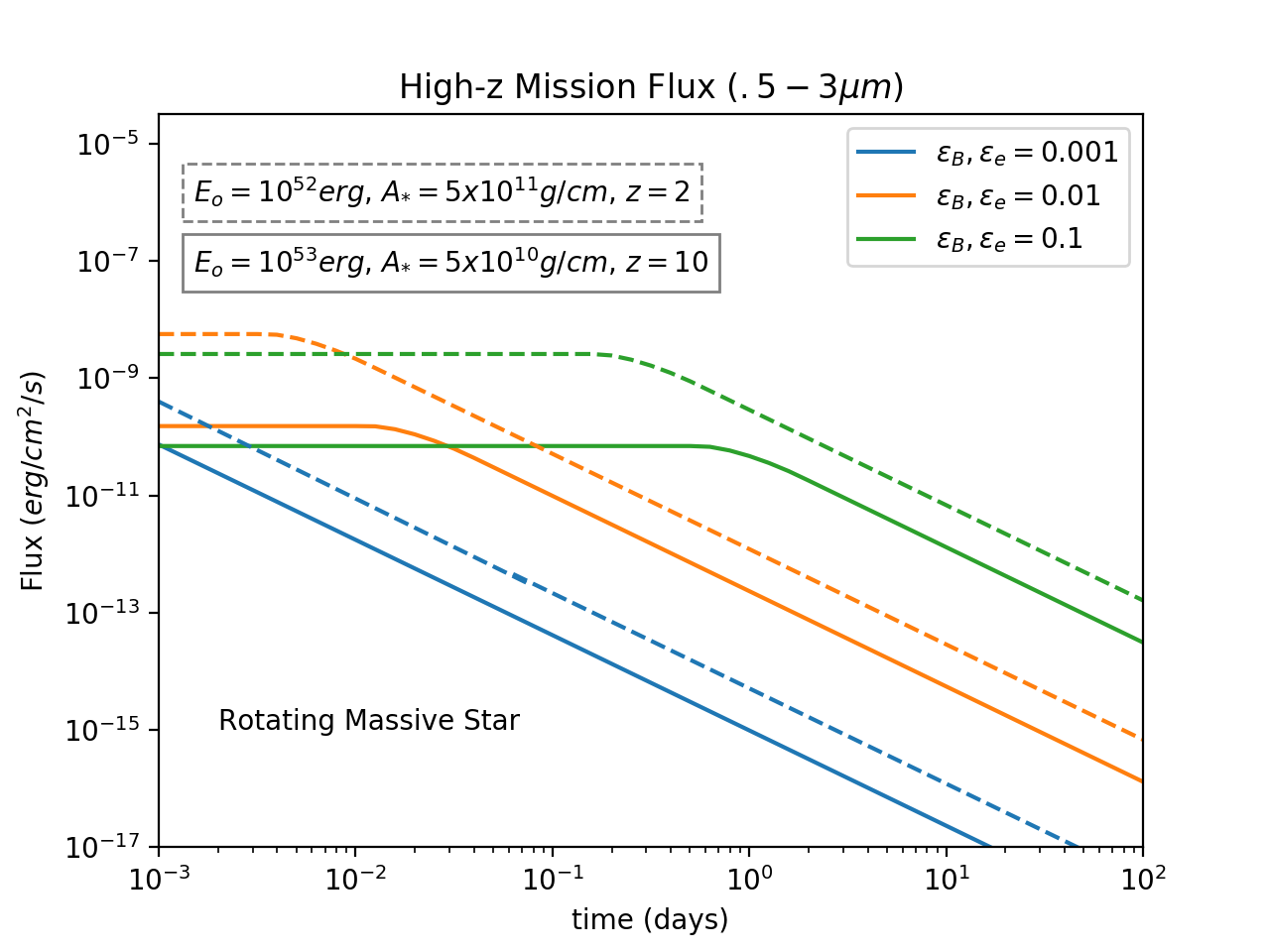}
\caption{Two fiducial lightcurves in the High-z bandpass for a rotating massive star models at a redshift of 10 (solid lines) and 2 (dashed lines) for different values of the fractions of energy in the electrons and magnetic field. The high redshift progenitor has less wind, but a higher isotropic emitted energy, while the lower redshift progenitor presumably has a stronger wind (due to higher metallicity) and a lower isotropic emitted energy.}
\label{fig:collapsar} 
\end{figure}

\begin{figure}
% Use the relevant command for your figure-insertion program
% to insert the figure file.
% For example, with the option graphics use
  \includegraphics[width=3.4in,height=3.5in]{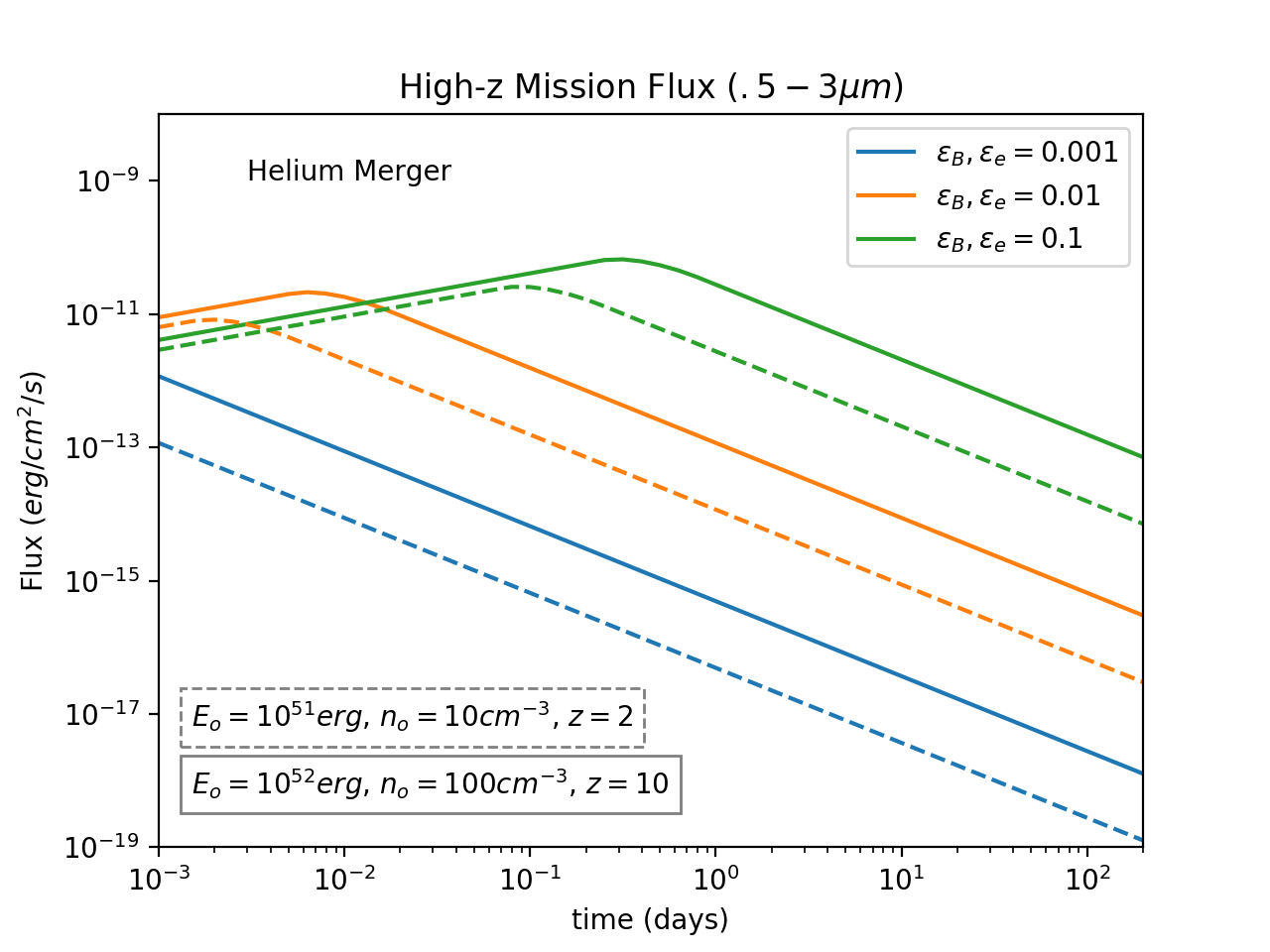}
\caption{Two fiducial lightcurves in the High-z bandpass for a Helium merger model at a redshift of 10 (solid lines) and 2 (dashed lines) for different values of the fractions of energy in the electrons and magnetic field. The high redshift progenitor has a higher isotropic emitted energy and a higher circumburst density, while the lower redshift progenitor has lower values of each of these quantities.}
\label{fig:hemerg} 
\end{figure}

\begin{figure}
% Use the relevant command for your figure-insertion program
% to insert the figure file.
% For example, with the option graphics use
 \includegraphics[width=3.4in,height=3.5in]{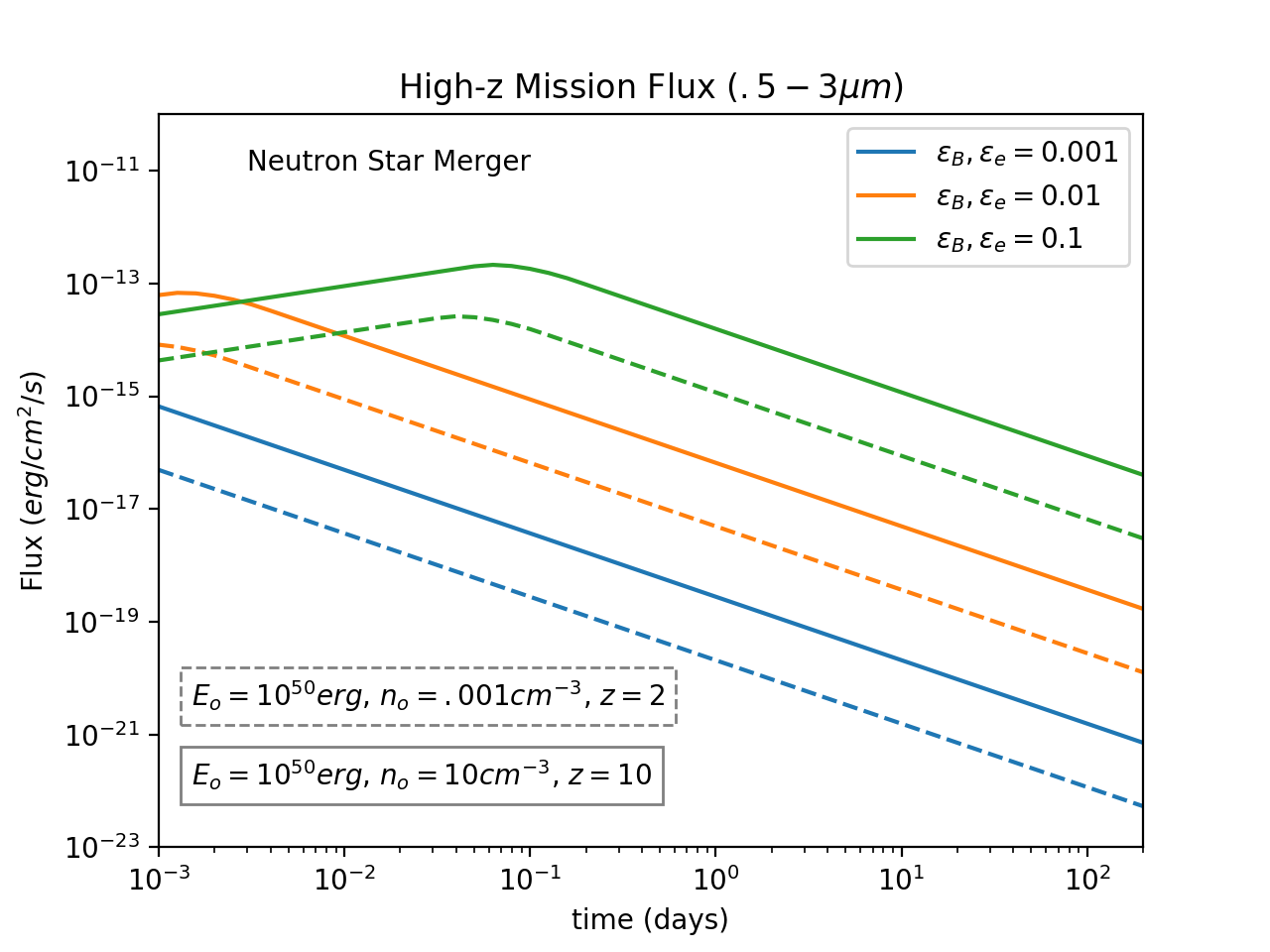}
\caption{Same as Figure 5, but using fiducial parameters for a Double Neutron Star merger progenitor at a redshift of 10 and 2.}
\label{fig:NSmerg} 
\end{figure}

\begin{figure*}
\centering
\includegraphics[width=3.5in,height=2.7in]{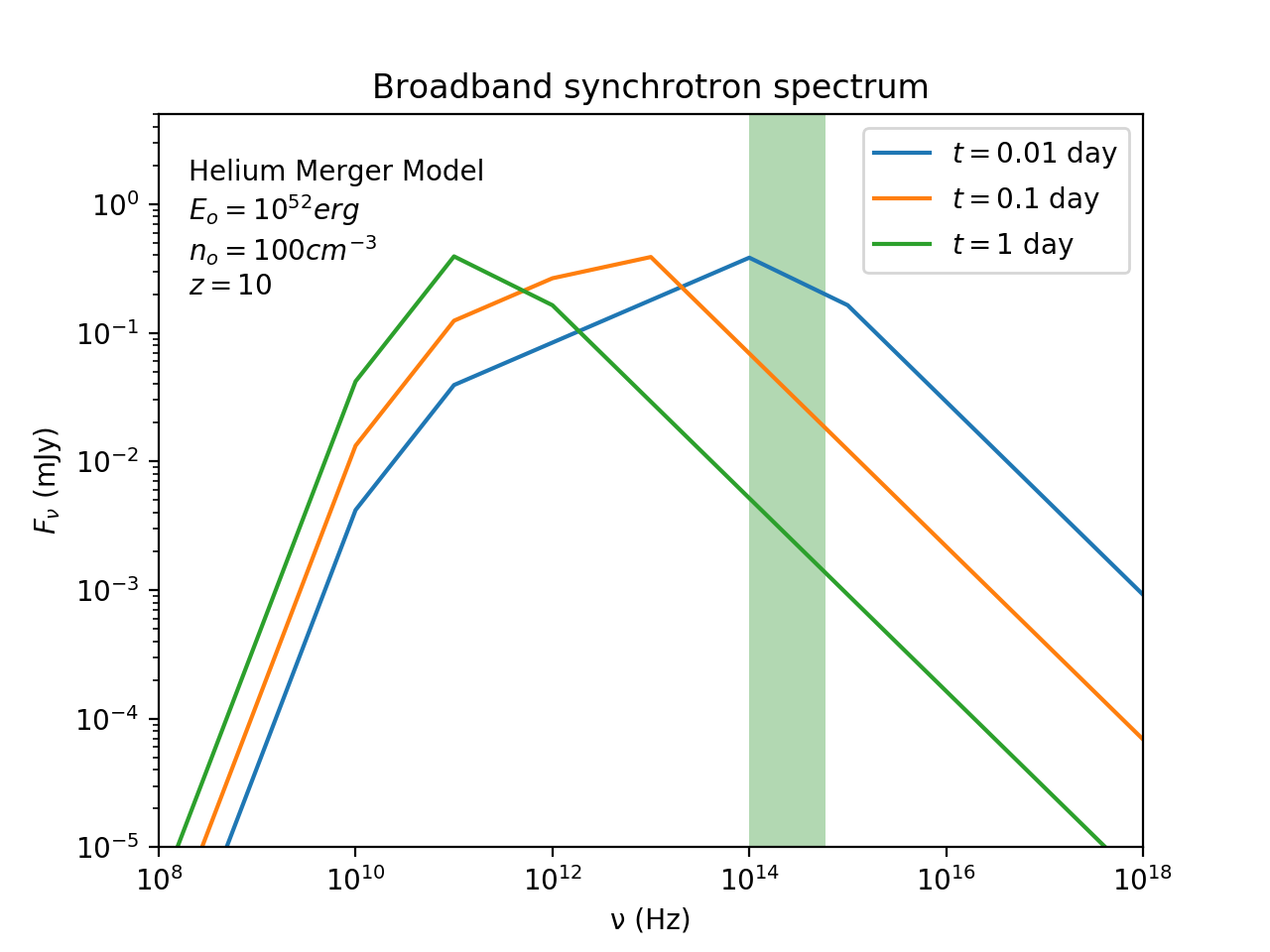}\includegraphics[width=3.5in,height=2.7in]{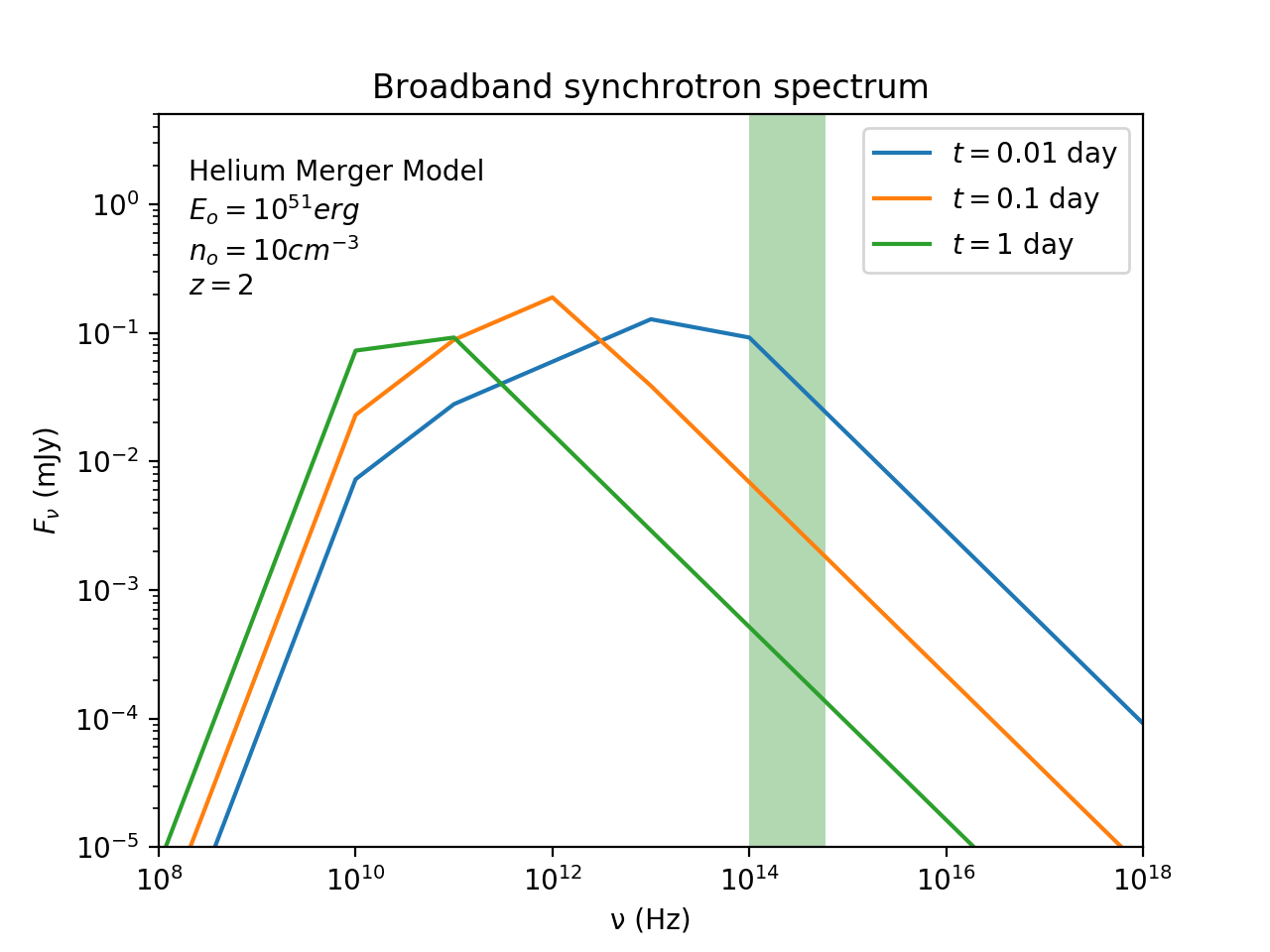}
\caption{Estimates of the broadband synchrotron spectra for the two Helium merger models shown in Figure 5.  The High-Z detector bandpass is highlighted}
\label{fig:spectra}
\end{figure*}

We can now apply the trends with metallicity in the initial mass function, stellar radii, and mass loss to the specific progenitors in this paper:
\begin{itemize}
\item{{\bf Massive Rotating Star}: The rate of this progenitor as a function
  of the star formation rate, especially those forming BHAD engines, will
  increase with redshift because the initial mass function produces
  more massive stars and mass loss is limited.  The cores of these
  stars might be slightly more compact, producing more powerful
  bursts, but this trend is not robust.  Similarly, there is no
  obvious trend in the duration.  The circumstellar medium (CSM) at 
  high redshift will have wind profiles with lower mass loss rates surrounded 
  by higher ambient densities (making smaller wind profile $r^{-2}$ regions), but if
  common envelope evolution drives the mass loss of the hydrogen
  envelope and helium winds are less sensitive to metals, the CSM properties 
  could be similar man not vary dramatically.}
\item{{\bf Tidal Locking}:  The flattening of the IMF and lower winds will 
increase the rate of this progenitor.  Although smaller radii might 
lower the rate, this scenario needs very close progenitors and is 
probably not strongly affected by the radius variations.  More fast-spinning 
cores are likely to be produced with the more compact stars, possibly making 
stronger/longer bursts.  Like all massive stars, the winds are weaker, but the 
ISM likely has higher densities (producing more compact wind bubbles.}
\item{{\bf He-He merger}:  Typically larger helium cores in this progenitor could 
produce more powerful and longer bursts.  These more massive helium cores will 
drive stronger winds within the possibly higher interstellar medium.}
\item{{\bf He-merger}:  The merger rate of a helium star with a NS or BH companion 
will increase for the same reasons as the rest of our massive star models.  Because 
the helium cores in these mergers will tend to be more massive, we expect these 
GRBs to be more powerful.  The specific angular momentum of these mergers will 
probably balance out and not make significantly longer bursts, but recall that 
this progenitor has been invoked already to make ultra-long bursts.  The CSM 
properties should mirror the rotating massive star models}
\item{{\bf NS/NS Mergers}:  The rate of NS mergers stays roughly constant, but with slightly higher NS masses, the disk masses are likely to be higher, producing stronger bursts.  More massive NSs are more compact, so the duration may be shorter.  \cite{wiggins18} found that higher redshift mergers occurred in higher densities (both because shorter period  binaries play a bigger role at higher redshift and because star-forming regions are  denser.  Figure~\ref{fig:nsmerg} shows the predicted densities for mergers as a function of redshift 
band.  This effect will be true for all of the compact merger events.}
\item{{\bf NS/BH mergers}:  Although the rate of NS/BH mergers will increase with redshift, larger black hole masses means that fewer of these systems will actually produce disks.  Otherwise  the trends in properies of this progenitor will follow NS/NS mergers.}
\item{{\bf WD/(NS,BH) mergers}:  The rate of these systems will increase with redshift.  More massive black holes will produce less powerful bursts but more massive white dwarfs will produce stronger bursts.  It is not clear what the final trend will be.  Similar effects will alter the duration.  The CSM is likely to be higher.}
\item{{\bf WD/WD Mergers:}  The flattening of the initial mass function will tend to lower 
the rate of these mergers, but it is possible that more massive white dwarfs 
will be produced that both make a GRB fate more likely and produce larger disks that drive more powerful, but shorter, bursts.  The CSM is likely to be higher.}
\end{itemize}
Table~\ref{tab:redshift} summarizes the variations in the model with
increasing redshift (decreasing metallicity) for the models accretion
disk models studied in this paper.  At this time, we can, at best,
discuss trends and only trends are listed in the table.  We show the
rate ($\Uparrow$ means an increase with respect to the star formation
rate, $\equiv$ means a similar value, and $\Downarrow$ means an
increase with respect to the star formation rate), power ($\Uparrow, \equiv, 
\Downarrow$ mean stronger, equivalent and weaker power respectively), duration 
($\Uparrow, \equiv, \Downarrow$ correspond to longer, shorter equivalent 
durations), and the CSM.  For the CSM, differences can occur both in the 
wind and interstellar medium (ISM) properties ($\Uparrow, \Downarrow$ correspond to higher, lower densities).  

\begin{table}
\caption{Variation with Increasing Redshift}
\label{tab:redshift}       % Give a unique label
% For LaTeX tables use
\begin{tabular}{lllll}
\hline\noalign{\smallskip}
Progenitor & Rate & Power & Dur. & CSM\\
\noalign{\smallskip}\hline\noalign{\smallskip}
Massive Stars & & & & \\
Rotating Star & $\Uparrow$ & $\Uparrow$? & $\equiv$ & $\Downarrow$? wind, $\Uparrow$? ISM \\
Tidal Locking & $\Uparrow$ & $\Uparrow$? & $\Uparrow$? & $\Downarrow$? wind, $\Uparrow$? ISM \\
He-He Merger & $\Uparrow$ & $\Uparrow$? & $\Uparrow$? & $\Uparrow$ wind, $\Uparrow$? ISM \\
He-Merger    & $\Uparrow$ & $\Uparrow$ & $\equiv$? & $\Downarrow$ wind, $\Uparrow$ ISM \\
\noalign{\smallskip}\hline
Compact& & & & \\
Mergers & & & & \\
NS/NS & $\equiv$ & $\Uparrow$? & $\Downarrow$? & $\Uparrow$ ISM \\
NS/BH  & $\Downarrow$ & $\Uparrow$ & $\Downarrow$ & $\Uparrow$ ISM \\
WD/(NS,BH)  & $\Uparrow$ & ? & ? & $\Uparrow$ ISM \\
WD/WD & $\Downarrow$ & $\Uparrow$? & $\Downarrow$? & $\Uparrow$ ISM \\

\noalign{\smallskip}\hline
\end{tabular}
% Or use
% \vspace*{5cm}  % with the correct table height
\end{table}

The trends in power and circumstellar medium will affect the emission from these bursts.  Figures~\ref{fig:collapsar}, ~\ref{fig:hemerg}, and ~\ref{fig:NSmerg} show the expected synchrotron light curves in the High-Z mission bandpass ($.5 - 3\mu m$) for three classes of models: rotating massive star, Helium merger, and NS-NS merger, respectively.  For each model, we show light curves at a redshift of 10 (solid lines) and 2 (dashed lines).  In addition, we plot the light curves for three different fractions of energy in the magnetic field and electrons (the three different colored lines).  Figure~\ref{fig:spectra} shows two sample broadband synchrotron spectra for the two Helium merger models employed in Figure~\ref{fig:hemerg}, and with the High-Z mission bandpass highlighted.

We note that although one might expect that the high redshift GRBs will exhibit a dimmer light curve due to their distances, the higher presumed emitted energy and circumburst density play against the redshift effect - in some cases the high redshift bursts are expected to have even more flux than low redshift bursts (see, for example, the Helium merger model case in Figure ~\ref{fig:hemerg}). Another way of saying this is that GRBs are not standard candles and their properties at high redshift can make them brighter. Indeed, \cite{lloydronning02} showed - using a large sample of GRBs with psuedo-redshifts - that GRBs exhibit strong luminosity evolution, with higher redshift burst being more luminous.
  
Broadband observations of high redshift GRBs will potentially allow us to put constraints on the emitted energy and circumburst density, which ultimately, we can tie back to the progenitor itself and thereby learn something about the metallicity and environments of these high redshift systems.  Combined with the more detailed, multi-messenger signals of nearby bursts, we can understand the physical mechanisms and underlying physics of GRBs and, in turn, use this understanding of GRBs to probe the early universe.

\section*{Acknowledgements}
This work was supported by the US Department of Energy through the Los Alamos National Laboratory. Additional funding was provided by the Laboratory Directed Research and Development Program under project number 20190021DR and the Center for Nonlinear Studies at Los Alamos National Laboratory under project number 20170508DR.

%
% BibTeX users please use
 \bibliographystyle{unsrtnat}
 \bibliography{ref.bib}

\end{document}